\documentclass[paper=A4,11pt,abstract=true,numbers=noenddot,
titlepage=false,twocolumn=false,
draft=false]{scrartcl}

\makeatletter
\DeclareOldFontCommand{\rm}{\normalfont\rmfamily}{\mathrm}
\DeclareOldFontCommand{\sf}{\normalfont\sffamily}{\mathsf}
\DeclareOldFontCommand{\tt}{\normalfont\ttfamily}{\mathtt}
\DeclareOldFontCommand{\bf}{\normalfont\bfseries}{\mathbf}
\DeclareOldFontCommand{\it}{\normalfont\itshape}{\mathit}
\DeclareOldFontCommand{\sl}{\normalfont\slshape}{\@nomath\sl}

\usepackage[usenames,dvipsnames]{color}
  \definecolor{hgreen}{rgb}{0,.3,0}
  \definecolor{hred}{rgb}{.3,0,0}
  \definecolor{hblue}{rgb}{0,0,.3}
  \definecolor{LightGray}{gray}{0.95}
\usepackage[
	    colorlinks=true,
	    linkcolor=hblue,
	    citecolor=hgreen,
	    filecolor=hblue,
	    urlcolor=hred
	    ]{hyperref}
\usepackage{pdfpages}
\usepackage{graphicx}
\usepackage{mathrsfs}
\usepackage{amsmath}
\usepackage{array}
\usepackage{amssymb}
\usepackage{dsfont}
\usepackage{braket}
\usepackage{bbold}
\usepackage{slashed}
\usepackage{cleveref}
\usepackage{subfig}
\usepackage[titletoc,title]{appendix}
\usepackage[affil-it]{authblk}
\usepackage{libertine}
\usepackage[numbers,sort&compress]{natbib}

\usepackage[numbers,sort&compress]{natbib}

\newcommand{\UV}{{\rm UV}}
\newcommand{\IR}{{\rm IR}}
\newcommand{\MS}{\overline{\rm MS}}
\newcommand{\BP}[3]{\langle #1\rangle^{(#2)}\bigl\vert_{#3}}
\newcommand{\Lag}{\mathscr{L}}
\newcommand{\Q}{\mathscr{Q}}
\newcommand{\E}{\mathscr{E}}
\newcommand{\N}{\mathscr{N}}

\newcommand{\PP}{\mathscr{P}}
\newcommand{\Z}{\mathcal{Z}}
\newcommand{\cO}{\mathscr{O}}
\newcommand{\BPsi}{\overline{\Psi}}
\newcommand{\sD}{\slashed{D}}
\newcommand{\sDr}{{\overset{\,\,\rightharpoonup}{\vphantom{a}\smash{\slashed{D}}}}}
\newcommand{\sDl}{{\overset{\,\,\leftharpoonup}{ \vphantom{a}\smash{\slashed{D}}}}}
\newcommand{\mira}{m_{\rm IRA}}

\newcommand{\oZ}[2]{\Z^{(#1)}_{#2}}
\newcommand{\dcZ}[2]{\Z^{(#1)}_{#2}}
\newcommand{\bdcZ}[2]{\fbox{$\Z^{{(#1)}}_{#2}$}}

\newcommand{\Nf}{N_{\!f}}
\newcommand{\order}[1]{ {\cal O}(#1)}

\newcommand{\Bil }[3]{B^{(#1)}_{\rm #2}} 
\newcommand{\cBil}[3]{\mathcal{B}^{(#1)}_{\rm #2}} 

\newcommand{\BilL }[3]{B^{(#1)}_{{\rm #2}~#3}}
\newcommand{\cBilL}[3]{\mathcal{B}^{(#1)}_{{\rm #2}~#3}}

\setcapindent{1em}
\setkomafont{captionlabel}{\bf}
\setkomafont{caption}{\it}
\numberwithin{equation}{section}

\begin{document}
\renewcommand\Authands{, }
\unitlength = 1mm

\title{Scaling dimensions in QED$\boldsymbol{_3}$\\from the $\boldsymbol{\epsilon}$-expansion }
\date{\today}
\author[a]{Lorenzo Di Pietro%
        \thanks{\texttt{ldipietro@perimeterinstitute.ca}}}
\author[b]{Emmanuel Stamou%
        \thanks{\texttt{estamou@uchicago.edu}}}
\affil[a]{Perimeter Institute for Theoretical Physics, Waterloo, ON\,N2L\,2Y5, Canada}
\affil[b]{Enrico Fermi Institute, University of Chicago, Chicago, IL 60637, USA}

\maketitle
\begin{abstract}

We study the fixed point that controls the IR dynamics of
QED in $d = 4 - 2\epsilon$.
We derive the scaling dimensions of four-fermion and bilinear operators beyond leading order
in $\epsilon$-expansion. 
For the four-fermion operators, this requires the computation of a two-loop 
mixing that was not known before.
We then extrapolate these scaling dimensions to $d = 3$ to estimate their value at the 
IR fixed point of QED$_3$ as function of the number of fermions $\Nf$. 
The next-to-leading order result for the four-fermion 
operators corrects significantly the leading one. 
Our best estimate at this order indicates that they do not cross marginality for any value of $\Nf$,
which would imply that they cannot trigger a departure from the conformal phase.
For the scaling dimensions of bilinear operators, we observe better convergence
as we increase the order.
In particular, $\epsilon$-expansion provides
a convincing estimate 
for the dimension of the flavor-singlet scalar  
in the full range of $\Nf$.
\end{abstract}
\pdfbookmark[1]{Table of Contents}{tableofcontents}
\setcounter{page}{1}
\tableofcontents

\section{Introduction\label{sec:introduction}}
Quantum Electrodynamics (QED) in $3d$ is an asymptotically-free gauge theory, 
which becomes strongly interacting in the IR. 
When the $U(1)$ gauge field is coupled to an even number, $2 \Nf$, of complex two-component 
fermions, and the Chern--Simons level is zero, the theory is parity invariant and has an 
$SU(2 \Nf)\times U(1)$ global symmetry.
For large $\Nf$ the theory flows in the IR to an interacting conformal field 
theory (CFT) that enjoys the same parity and global symmetry.
The CFT observables are then amenable to perturbation theory in $1/\Nf$;
this has been done for scaling dimensions \cite{Gracey:1993sn, Gracey:1993iu, PhysRevB.66.144501, PhysRevB.72.104404, 
PhysRevB.76.149906, PhysRevB.77.155105, PhysRevB.78.054432, Borokhov:2002ib, Pufu:2013vpa, 
Dyer:2013fja, Chester:2016ref}, two-point functions 
of conserved currents \cite{Huh:2014eea, Huh:2013vga, Giombi:2016fct}, and the free energy 
\cite{Klebanov:2011td}.
The IR fixed point is expected to persist beyond this large-$\Nf$ regime, but
not much is known about it. 
Ref.~\cite{Chester:2016wrc} employed  the  conformal bootstrap approach to derive bounds 
on the scaling dimensions of some monopole operators. 
Another method to study the small-$\Nf$ CFT 
is the $\epsilon$-expansion, which
exploits the existence of a fixed point of 
Wilson--Fisher-type \cite{Wilson:1971dc} 
in QED continued to $d = 4 - 2\epsilon$ dimensions.
When $\epsilon \ll 1$ we can access observables
via a perturbative expansion in $\epsilon$ and subsequently 
attempt  an extrapolation to $\epsilon = \frac{1}{2}$. 
The $\epsilon$-expansion of QED was employed to
estimate some scaling dimensions
\cite{DiPietro:2015taa, Chester:2015wao}, 
the free energy $F$ \cite{Giombi:2015haa}, 
and the coefficients $C_T$ and $C_J$ \cite{Giombi:2016fct}.
In particular, ref.~\cite{DiPietro:2015taa} considered operators made out 
of gauge-invariant products of either four or two fermion fields. 

Four-fermion operators are interesting because of the
dynamical role they can play in the transition from the 
conformal to a symmetry-breaking phase, which is conjectured to 
exist  if $\Nf$ is smaller than a certain critical number $\Nf^c$ 
\cite{Pisarski:1984dj, Vafa:1984xh, Appelquist:1988sr, Appelquist:2004ib}. 
In fact, the operators with the lowest UV dimension that are  singlet under 
the symmetries of the theory are four-fermion operators.
If for small $\Nf$ they are {\it dangerously irrelevant}, i.e., their anomalous dimension 
is large enough for them to flow to relevant operators in the IR, they may trigger 
the aforementioned transition \cite{Kaveh:2004qa, PhysRevB.78.054432, Braun:2014wja}.\footnote{More precisely, when the four-fermion operators are slightly irrelevant in the IR, there is an additional nearby UV fixed point. As these operators become marginal, the two fixed points cross each other, and they can annihilate and disappear \cite{Kaplan:2009kr}. For a more detailed discussion, see section 5 of ref. \cite{Giombi:2015haa}. Ref. \cite{Gukov:2016tnp} pointed out that in order to describe properly the conjectured transition, one cannot ignore higher-order terms in the four-fermion couplings. A study of the RG flow that employed $\epsilon$-expansion and included four-fermion couplings appeared recently in ref. \cite{Janssen:2017eeu}.
}
The one-loop result of ref.~\cite{DiPietro:2015taa} led to the  estimate $\Nf^c \leq 2$. 
 
Bilinear operators, i.e., operators with two fermion fields, are interesting
because they are presumably among the operators with lowest dimension.
For instance, when continued to $d = 3$, the two-form operators
$\BPsi \gamma_{[\mu}\gamma_{\nu]} \Psi$ become the additional conserved currents 
of the $SU(2 \Nf)$ symmetry, of which only a $SU(\Nf)$ subgroup is visible in $d = 4-2\epsilon$. 
This leads to the conjecture that their scaling dimension should approach the value 
$\Delta = 2$ as $\epsilon \to \frac{1}{2}$, which was tested at the one-loop
level  in ref.~\cite{DiPietro:2015taa}.
 
In order to assess the reliability of the $\epsilon$-expansion in QED, and  
improve the estimates from the one-loop extrapolations, 
it is desirable to extend the calculation of these anomalous dimensions beyond 
leading order in $\epsilon$. This is the purpose of the present paper. 
Let us describe the computations we perform and the significance of the
results. 
 
We first consider four-fermion operators. In the UV theory in 
$d=4 - 2 \epsilon$, there are two such operators that upon continuation 
to $d = 3$ match with the singlets of the $SU(2 \Nf)$ symmetry. We compute their anomalous 
dimension matrix (ADM) at two-loop level by renormalizing off-shell, amputated Green's functions 
of elementary fields with a single operator insertion. 
As we discuss in detail in a companion paper \cite{DiPietroStamou1}, 
knowing this two-by-two ADM is not sufficient to obtain the $\order{\epsilon^2}$ scaling dimensions 
at the IR fixed point. 
We also need to take into account the full one-loop mixing with a family of infinitely many  
operators that have the same dimension in the free theory. 
These operators are of the form
\begin{equation}
(\BPsi \Gamma^n_{\mu_1\dots \mu_n}\Psi )^2~,
\end{equation}
where $n$ is an odd integer, and $\Gamma^{n}_{\mu_1\dots \mu_n}\equiv \gamma_{[\mu_1}\dots \gamma_{\mu_n]}$ 
is an antisymmetrized product of gamma matrices. 
All the operators in this family except for the first two, i.e., $n=1,3$, vanish for the integer 
values $d=4$ and $d=3$, but are non-trivial for intermediate values $3 < d < 4$. For this reason 
they are called {\it evanescent} operators. Taking properly into account the 
contribution of the evanescent operators, via the approach described in ref.~\cite{DiPietroStamou1}, 
we obtain the next-to-leading order (NLO) scaling dimension of the first two operators. 
We then extrapolate to $\epsilon = \frac {1}{2}$ using a Pad\'e approximant, leading to the result 
presented in subsection~\ref{sec:4fermiond3} and summarized in figure~\ref{fig:DeltaNf}. 
The deviation from the leading order (LO) scaling dimension 
is considerable for small $\Nf$, indicating that at this order we cannot yet
obtain a precise estimate for this observable of the three-dimensional CFT.
Taking, however, the NLO result at face value, we would conclude that the 
four-fermion operators are never dangerously irrelevant. 
This resonates with recent results that suggest that
QED$_3$ is conformal in the IR for any value of $\Nf$.
Namely, refs.~\cite{Karch:2016aux, Hsin:2016blu, Benini:2017dus}
argued based on $3d$ bosonization dualities 
\cite{Aharony:2015mjs, Karch:2016sxi, Murugan:2016zal, Seiberg:2016gmd}, 
that for $\Nf = 1$ the $SU(2)\times U(1)$ symmetry is in fact
enhanced to $O(4)$ (this is related to the self-duality present
in this theory \cite{Xu:2015lxa}).
Also, a recent lattice study \cite{Karthik:2015sgq} found no evidence for
a symmetry-breaking condensate (for previous lattice studies see 
refs.~\cite{Hands:2002qt, Hands:2004bh, Strouthos:2008hs}).
 
We then consider the bilinear ``tensor-current'' operators of the 
form
\begin{equation}
\BPsi \Gamma^n_{\mu_1 \dots \mu_n} \Psi~,
\end{equation} 
for $n=0,1,2,3$. 
We obtain their IR scaling dimension up to $\order{\epsilon^3}$ using the
three-loop computations from ref.~\cite{Gracey:2000am}. 
Having these higher-order results, we are in the position to 
employ  different Pad\'e approximants to  estimate errors 
and test the convergence as we increase 
the order.
As mentioned above, in the limit $d\to 3$ the operators with $n=1,2$ 
approach conserved currents of the $SU(2\Nf)$ symmetry. 
Indeed, we show in subsection~\ref{sec:bilinearsd3} 
(see figure \ref{fig:DeltaBilinearsNftest}) that the extrapolated 
scaling dimension of the two-form operators approaches the value 
$\Delta = 2$ as we increase the order. 
As $d\to 3$, the operators with $n = 0,3$   
approach scalar bilinears, which are either in the 
adjoint representation of $SU(2\Nf)$ or are singlets. 
For the singlet scalar, which  is continued by a bilinear with $n = 3$,
the results of various extrapolations we perform are all close 
to each other (see figure~\ref{fig:DeltaBilinearsNfpredictions}), 
indicating that $\epsilon$-expansion provides
a good estimate for this scaling dimension 
in the full range of $\Nf$. 
For the adjoint scalar, different
components are continued by operators with either $n=0$ 
or $n=3$, giving two independent extrapolations at each order 
in $\epsilon$. 
As expected, we find  that the two independent extrapolations 
approach each other as we increase the order 
(see figure~\ref{fig:DeltaBilinearsNfpredictions}).

The rest of the paper is organized as follows: in section 
\ref{sec:QEDind4m2ep} we set up our notation and describe
the fixed point of QED in $d=4-2\epsilon$; 
in section \ref{sec:4fermionind4m2ep} we present 
the computation of the two-loop  ADM of the four-fermion operators, 
and then the result for their scaling dimension at the IR fixed point 
in $d=4-2\epsilon$; 
in section \ref{sec:bilinearsind4m2ep} we present the same result for the 
bilinear operators; 
in section \ref{sec:d3extrapolation} we extrapolate the scaling dimensions to $d=3$, 
and plot the resulting
dimensions as a function of $\Nf$ for the various operators 
we consider; finally in section \ref{sec:conclusions} we present our
conclusions and discuss possible future directions. In the appendices we collect
additional material and some useful intermediate results.

\section{\texorpdfstring{QED in $\boldsymbol{d=4-2\epsilon}$\label{sec:QEDind4m2ep}}{%
	 QED in d=4-2ep}}

We consider QED with $\Nf$ Dirac fermions $\Psi^a$, $a=1,\dots,\Nf$, of charge $1$. 
The Lagrangian is
\begin{equation}
\Lag_{\text{QED}}  = - {1\over 4} F^{\mu\nu} F_{\mu\nu} +  \BPsi_a i \gamma^\mu D_\mu \Psi^a~,\label{eq:Lag}
\end{equation}
with the  covariant derivative defined as
\begin{equation}
D_\mu  \equiv \partial_\mu + i  e A_\mu \,.
\end{equation}
Summation over repeated flavor indices is implicit. 
We work in the $R_\xi$-gauge, defined by 
adding the gauge-fixing term
\begin{equation}
\Lag_{\text{g.f.}} = - {1 \over 2 \xi} (\partial_\mu A^\mu)^2~.
\end{equation}
We collect the Feynman rules in appendix~\ref{app:feynmanrules}.

The algebra of the gamma matrices is $\{\gamma_\mu, \gamma_\nu\} = 2 \eta_{\mu\nu}$, 
with $\eta_{\mu\nu}\eta^{\nu\rho} = \delta_\mu^\rho$ and $\delta^\mu_\mu = d$.
We will employ some useful results on $d$-dimensional 
Clifford algebras from ref.~\cite{Kennedy:1981kp}. 
We normalize the traces by ${\rm Tr}[\mathbb{1}] = 4$, for any $d$. For $d=3$, 
$\Psi^a$ decomposes as
\begin{equation}
\label{eq:Diracd3}
\Psi^a \underset{d \to 3}{\longrightarrow} \begin{bmatrix}
			\psi^a \\
			\psi^{a+\Nf}
                     \end{bmatrix} ~,
\end{equation}
giving $2\Nf$ complex two-component 
$3d$ fermions $\psi^i$, $i=1,\dots,2\Nf$, all with charge $1$. 
Correspondingly, the gamma matrices decompose as
\begin{equation}
\label{eq:gammad3}
\gamma_\mu \underset{d \to 3}{\longrightarrow} \begin{bmatrix}
			0 & \gamma^{(3)}_\mu   \\
			\gamma^{(3)}_\mu & 0
                     \end{bmatrix} ~,
\end{equation}
where $\{\gamma_\mu^{(3)}\}_{\mu=1,2,3}$ are two-by-two $3d$ gamma matrices.

In $d=4$, the global symmetry preserved by the gauge-coupling is 
$SU(\Nf)_L\times SU(\Nf)_R$.
In $d = 4-2\epsilon$, evanescent operators violate the 
conservation of the nonsinglet axial currents \cite{Collins:1984xc}, 
so only the diagonal subgroup $SU(\Nf)$ is preserved.
In $d=3$, this symmetry enhances to $SU(2 \Nf) \times U(1)$.

We define $\alpha\equiv\frac{e^2}{16\pi^2}$ and  denote 
bare quantities with a subscript ``$0$''. The renormalized
coupling is given by
\begin{equation}
\alpha_0 = Z_\alpha \alpha(\mu) \mu^{2 \epsilon}~,
\end{equation}
where the renormalization constant $Z_\alpha(\alpha,\epsilon)$ absorbs the
poles at $\epsilon=0$, and $\mu$ is the renormalization scale. 
The beta function reads 
\begin{align}
\label{eq:alpharun}
\frac{d \alpha }{d\log\mu}  = -2 \epsilon \alpha +  \beta(\alpha, \epsilon)~,
\end{align}
where
\begin{align}
\beta(\alpha, \epsilon) \equiv  
-\alpha \frac{d \,\log Z_\alpha}{d \log\mu}~.
\end{align}
In Minimal Subtraction ($\MS$), $\beta$ depends only on $\alpha$ 
and not on $\epsilon$.
The $\MS$ QED $\beta$ function is known up to four-loop order 
for generic $\Nf$ \cite{Gorishnii:1987fy, Gorishnii:1990kd}
\begin{multline}
\beta(\alpha) = 
 \frac{8}{3} \Nf \alpha^2 
+ 8 \Nf \alpha^3 
- \left(\frac{88}{9} \Nf^2+4 \Nf \right)\alpha^4 \\
- \left(\frac{2464}{243} \Nf^3+\frac{16}{27} \left(312 \zeta(3)-95\right) \Nf^2+92 \Nf\right)\alpha^5
+ \order{\alpha^6}~.\label{eq:twoloopbeta}
\end{multline}
Using eqs.~\eqref{eq:alpharun} and \eqref{eq:twoloopbeta} we find 
that in $d=4-2\epsilon$ the theory has a fixed point at
\begin{multline}
\label{eq:alphaStar}
\alpha^* = \frac{3}{4 \Nf}\epsilon - \frac{27}{16 \Nf^2}\epsilon^2 + 
\frac{9(22 \Nf+117)}{128\Nf^3} \epsilon^3 \\
+\frac{\left(308 \Nf^2+9 (624 \zeta(3)-685)\Nf -9963\right)}{256 \Nf^4}
\epsilon^4
+ \order{\epsilon^5}\,,
\end{multline}
with $\zeta(n)$ the Riemann zeta function.

Our convention for renormalizing fields is  
\begin{align}
\Psi^{a}_0 =  Z_\Psi^{1/2} \Psi^a,\quad  A^{\mu}_0 = Z_A^{1/2} A^\mu\,.
\label{}
\end{align}
By the Ward Identity, $Z_A = Z_\alpha^{-1}$.
For our computations we need the field-renormalization of the fermion
up to two-loop order. In $\MS$ and generic $R_\xi$-gauge
it reads
\begin{equation}
Z_\Psi = 1- \frac{\alpha}{\epsilon}\xi + \frac{\alpha^2}{\epsilon^2}\xi^2 + \frac{\alpha^2}{\epsilon}\left(\frac 34 + \Nf\right)+\order{\alpha^3}~.
\end{equation}

\subsection{Operator mixing\label{sec:operatormixing}}
To compute the anomalous dimension of local operators $\cO^i$, 
we add these operators to the Lagrangian
\begin{equation}
\Lag_{{\text{QED}}} \to \Lag_{{\text{QED}}} + \sum_i (C_0)^i (\cO_0)_i ~,
\end{equation} 
and compute their renormalized couplings $C^i$ at linear level 
in the bare ones
\begin{equation}
(C_0)^j = C^i  \Z_i^{~j}~.
\end{equation}
$\Z_i^{~j}$ are the mixing renormalization constants from which we obtain 
the ADM
\begin{equation}
{\gamma(\alpha, \epsilon)} = - \frac{d \log \Z }{d \log\mu} ~.
\end{equation}
Like $\beta$, $\gamma$ does not depend on $\epsilon$ in the $\MS$ scheme.
We introduce the following notation for the coefficients of the expansion
in $\alpha$ and $\epsilon$
\begin{equation}
\label{eq:Zexpansion}
\Z(\alpha, \epsilon) = \mathbb{1} + \sum_{L=1}^{\infty}\alpha^L \sum_{M=-L}^{\infty} \epsilon^{M} \oZ{L,-M}{} ~.
\end{equation}

The most direct way to compute the mixing $\Z_i^{~j}$ is to renormalize
amputated one-particle-irreducible Green's functions
with zero-momentum operator insertions and elementary fields
as external legs.
Alternatively, one can renormalize the
two-point functions of the composite operators.
The former method has two main advantages.
The first is that to extract $n$-loop poles only $n$-loop diagrams need to be computed.
The second is that we can insert the operators with zero-momentum.
This makes higher-loop computations more tractable.
The disadvantage is that off-shell Green's functions 
with elementary fields as external legs are not gauge-invariant,
so some results in the intermediate steps of 
the calculation are $\xi$-dependent, which is why we need to 
include the $\xi$-dependent wave-function renormalization 
of external fermions.
In addition, operators that vanish under the equations of 
motion (EOM) enter the renormalization of such off-shell Green's functions.
We refer to the latter as EOM-vanishing operators. 

In the next section, we  consider composite operators given by 
scalar quadrilinear and bilinear operators in the fermion fields. 
We first present the computation of the two-loop 
anomalous dimension of the four-fermion operators and use 
it to obtain the $\order{\epsilon^2}$ IR scaling dimension 
at the fixed point. 
Next, we employ the already existing results of the three-loop 
anomalous dimension of bilinear operators \cite{Gracey:2000am} 
to obtain their IR dimension to $\order{\epsilon^3}$. 

\section{\texorpdfstring{Four-fermion operators in $\boldsymbol{d=4-2\epsilon}$\label{sec:4fermionind4m2ep}}{%
	Four-fermion operators in d=4-2ep}}

In this section, we present the computation of the ADM
of the four-fermion operators
\begin{equation}
\begin{split}
\Q_1  &= (\BPsi_a \gamma_\mu \Psi^a)^2, \\
\Q_3  &=(\BPsi_a\Gamma^{3}_{\mu_1\mu_2\mu_3}\Psi^a)^2\,,
\end{split}
\label{eq:Q1Q3}
\end{equation}
at the two-loop level. 
Here and in the following 
$\Gamma^{n}_{\mu_1\dots \mu_n}\equiv \gamma_{[\mu_1}\dots \gamma_{\mu_n]}$, 
with the square brackets denoting antisymmetrization, which includes the 
conventional normalization factor $\frac{1}{n!}$.

In $d=4$, the operators in eq.~\eqref{eq:Q1Q3} are the only two operators 
with scaling dimension $6$ at the free fixed point that are singlets under
the global symmetry $SU(\Nf)_L\times SU(\Nf)_R$. 
We focus on these flavor-singlet operators because, as explained in 
the introduction, we are interested in understanding
whether or not they are relevant at the IR fixed point. 
The calculation of the ADM for flavor-nonsinglet
operators is actually simpler because it involves a subset 
of the diagrams. 
We report the result for some nonsinglet operators in 
appendix~\ref{app:offdiagonal}.

In $d = 4-2\epsilon$, insertions of $\Q_1$ and $\Q_3$ 
in loop diagrams generate additional structures
that are linearly independent to the Feynman rules of $\Q_1$ and $\Q_3$.
To renormalize the divergences proportional to such structures, 
we need to enlarge the operator basis. 
It is most convenient to define the complete basis 
by adding operators that vanish for $\epsilon\to 0$, 
and hence are called {\it evanescent} operators,
as opposed to $\Q_1$ and $\Q_3$ that we refer to as
{\it physical} operators.
There is an infinite set of such evanescent operators.
One choice of basis for them is
\begin{equation}
\E_n  = (\BPsi_a\Gamma^n_{\mu_1\dots \mu_n}\Psi^a)^2
         + \epsilon a_n \Q_1 + \epsilon b_n \Q_3\,,
\label{eq:evadef}
\end{equation}
with $n$ an odd integer $\ge 5$. The terms proportional to the arbitrary
constants $a_{n}$ and $b_{n}$ are of the form $\epsilon \times$ a physical operator; 
they parametrize different possible choices for the basis of
evanescent operators.

For the computation of the ADM we adopt the subtraction 
scheme introduced in refs.~\cite{Dugan:1990df, Bondi:1988fp}. 
Since this is the most commonly used scheme for applications
in flavor physics, we refer to it as the {\it flavor scheme}. 
We label indices of the ADM using odd integers $n\leq 1$, 
so that $n=1,3$ correspond to the physical operators, eq.~\eqref{eq:Q1Q3},
and $n\geq 5$ to the evanescent operators, eq.~\eqref{eq:evadef}. 
The ADM up to two-loop order is\footnote{The one-loop ($\Q_1$, $\Q_3$) 
block of the ADM can be found in ref.~\cite{Beneke:1995qq}; 
it is  sufficient to obtain the $\order{\epsilon}$ prediction
of the scaling dimensions \cite{DiPietro:2015taa}.}
\begin{equation}
\label{eq:anodimalphaff}
\gamma(\alpha, \epsilon) = \alpha \left(\gamma^{(1,0)}+\epsilon\, \gamma^{(1,-1)} + \order{\epsilon^2}\right) + \alpha^2\left(\gamma^{(2,0)} + \order{\epsilon}\right)~,
\end{equation}
where
\begin{align}
\gamma^{(1,0)}_{n m} &=
\left\{
\begin{array}[]{cl}
16\delta_{n3}+ 2n(n-1)(n-5)(n-6)&\text{for}\quad m=n-2~,\\
\frac{8}{3}(2\Nf+1)\delta_{n1} -4(n-1)(n-3)	&\text{for}\quad m=n~,\\
2 		&\text{for}\quad m=n+2~,\\
0 		&\text{otherwise\,,}\\
\end{array}
\right. \label{eq:gamma10}\\[1em]
\gamma^{(1,-1)}_{nm} &=
\left\{
\begin{array}[]{ll}
	 32 (-1)^{\frac{n(n-1)}{2}} (n-2)(n-5)!\\
	\quad- 2  n(n-1)(n-5)(n-6) a_{n-2}\\
	\qquad+ \left(\frac{8}{3}(2\Nf+1)+ 4 (n-1)(n-3)\right) a_{n}\\
	\qquad\quad- 2 a_{n+2} + 88 b_n
	&\text{for}~m=1,\,n\ge5~,\\[1em]
	- 80 \delta_{n5}\\
	\quad-  2n(n-1)(n-5)(n-6)  b_{n-2}\\
	\qquad+ 4(n-1)(n-3) b_{n}- 2b_{n+2}+ 2a_n 
	&\text{for}~m=3,\,n\ge5~,\\[1em]
\hspace*{7em}0 &\text{otherwise\,,}
\end{array}
\right.  \label{eq:gamma1min1}\\[1em]
\gamma^{(2,0)}_{nm} &=
\left\{
\begin{array}[]{ll}
	\begin{bmatrix}
          -  \frac{2}{27} (2275+8 \Nf)    & -\frac{4}{9}  (49+ 3 \Nf) \\
             \frac{16}{9} (199 + 107 \Nf) & 110+ \frac{80 \Nf}{3}
	\end{bmatrix} +&\\
	+ a_5 
	\begin{bmatrix}
 		-2 & 0 \\
 		\frac{8}{3} (1+\Nf) & 2 \\
	\end{bmatrix}
 	+ b_5 
	\begin{bmatrix}
 		0 & -2 \\
 		88 & -\frac{8}{3} \Nf \\
	\end{bmatrix}
	&\text{for}~n,m=1,3~,\\[2em]
\hspace*{7em}0	&\text{for}~n\ge5~\text{and}~m=1,3\,,\\[1em]
\hspace*{5em}\text{not required}  &\text{otherwise}\,.
\end{array}
\right.\label{eq:g20}
\end{align}

At the two-loop level we have shown only the entries in the
physical--physical
two-by-two $(\Q_1, \Q_3)$ block,
and the evanescent--physical entries,
which by construction vanish in the flavor scheme.
No other two-loop entry enters the prediction of the $\order{\epsilon^2}$
prediction of the scaling dimensions at the fixed point.

Notice that the invariant $(\Q_1,\Q_3)$ block of $\gamma^{(2,0)}$ depends on the 
coefficients $a_5$ and $b_5$, which parametrize our choice of basis. 
This dependence can be understood as a sign of 
scheme-dependence \cite{Herrlich:1994kh}.
Clearly, this implies that the scaling dimensions at $\order{\epsilon^2}$ 
are not simply obtained from the eigenvalues 
of this invariant block, as also its eigenvalues depend on $a_5$ and $b_5$. 
The additional contribution that cancels this basis-dependence 
originates from the $\order{\epsilon}$ term $\gamma^{(1,-1)}$
in the one-loop ADM. 
Such $\order{\epsilon}$ terms are indeed  induced in every scheme 
that contains finite renormalizations, such as the flavor scheme. 
For a thorough discussion of the scheme/basis-dependence and 
its cancellation we refer to ref.~\cite{DiPietroStamou1}.

There are a few non-trivial ways of partially 
testing the correctness of the two-loop results:
\begin{itemize}
\item[{\it i)}]{We performed all computations in general $R_\xi$ gauge.
This allowed us to explicitly check that the mixing of
 gauge-invariant operators indeed does not depend on 
$\xi$.}
\item[{\it ii)}]{All the two-loop counterterms are local, i.e., 
the local counterterms from one-loop diagrams subtract
all terms proportional to $\frac{1}{\epsilon} \log\mu$
in two-loop diagrams.
}
\item[{\it iii)}]{The $\frac{1}{\epsilon^2}$ poles of the 
two-loop mixing constants satisfy the relation
\begin{align}
\oZ{2,2}{} = \frac 12 \oZ{1,1}{}\oZ{1,1}{} - \frac{1}{2} \beta^{(1,0)}\oZ{1,1}{}\,,
\end{align}
where $\beta^{(1,0)}$ is the one-loop coefficient of the beta-function. This is equivalent to the $\epsilon$-independence of the anomalous dimension
\cite{Faddeev:1967fc}.}
\end{itemize}

In the next two subsections, we discuss the renormalization 
of the one- and two-loop Green's functions from which we extract 
the relevant entries of the mixing matrix $\Z$ ---and 
ultimately the ADM entries in eqs.~\eqref{eq:gamma10}, \eqref{eq:gamma1min1}, and 
\eqref{eq:g20}--- 
and some technical aspects 
of the two-loop computation. 
A reader more interested in the results for the scaling dimensions 
may proceed directly to section~\ref{sec:evanescent}.

\subsection{Operator basis\label{sec:operatorbasis}}

As argued in section \ref{sec:operatormixing}, in general 
we need to consider also EOM-vanishing operators when renormalizing 
off-shell Green's functions. 
Moreover, in our computation we adopt an IR regulator 
that breaks gauge-invariance, so we also need to take into 
account some gauge-variant operators.
Below we list all operators that,
together with $(\Q_1,\Q_3)$ and $\{\E_n\}_{n\geq 5}$, 
enter the renormalization of the two-loop Green's functions we consider:

\paragraph{EOM-vanishing operators}
There is a single EOM-vanishing operator, $\N_1$, that affects
the ADM at the one-loop level and another one, $\N_2$,
that affects it at the two-loop level.
They read
\begin{align}
\N_1&=
        \frac{1}{e}\partial^\nu F_{\mu\nu} (\BPsi_a \gamma^\mu \Psi^a) +  \Q_1\equiv \N_1^\gamma+\Q_1\,,\label{eq:defN1}\\
\N_2&=
	\frac{1}{e} \partial^\nu F_{\mu\nu} (\BPsi_a \gamma^\mu \Psi^a)+
        \frac{1}{e^2}(\partial_\rho F^{\mu\rho})(\partial^\nu F_{\mu\nu}) \equiv  \N_1^\gamma +\N_2^{\gamma\gamma}~.
\end{align}
Additionally, there are EOM-vanishing operators 
that are only necessary to close the basis of independent 
Lorentz structures for certain Green's functions.
For completeness, we list them here
\begin{align}
\N_3 &=
        i \BPsi_a \sDr\sDr\sDr \Psi^a\,,\\
\N_4 &=
       \BPsi_a(
	\sDl\gamma^\mu\gamma^\nu+ 
	\gamma^\mu\gamma^\nu \sDr )
	\Psi^a F_{\mu\nu}~.
\end{align}
Here $\sD \equiv \gamma^\mu D_\mu$ and the arrow indicates on 
which field the derivative is acting.

\paragraph{Gauge-variant operators}
Renormalization constants subtract UV poles of Green's functions.
It is thus essential to ensure that no IR poles are mistakenly included in the
renormalization constants.
In practice, this means that an energy scale must be present
in dimensionally regularized integrals.
Otherwise, UV and IR contributions cancel each other 
and the result of the loop integral is zero in dimensional 
regularization \cite{Collins:1984xc}.

One possibility to introduce a scale is to keep the
external momentum in the loop integral.
However, {\it i)} such loop integrals are more involved 
than integrals obtained by expanding in powers of external 
momenta over loop momenta, and {\it ii)} keeping external 
momenta does not necessarily cure all the IR divergences, 
e.g., diagrams with gluonic snails in non-abelian gauge theories.
Another possibility for QED would be to introduce a mass for the Dirac fermions.
The drawback in this case is that we
would have to consider many more EOM-vanishing operators.

Instead, we apply the method of ``Infrared Rearrangement''
\cite{Misiak:1994zw,Chetyrkin:1997fm}.
This method consists in rewriting the massless propagators as a sum
of a term with a reduced degree of divergence and a term depending
on an artificial mass, $\mira$. 
Section~\ref{sec:Loops} contains some more details about the method.
The caveat is that the method violates gauge invariance in intermediate
steps of the computation.  
All breaking of gauge invariance is
proportional to $\mira^2$ and explicitly
cancels in physical quantities. 
However, to restore  gauge-invariance,
also gauge-variant  operators proportional to $\mira^2$ need to 
be consistently included in the computation.
Fortunately, due to the factor of $\mira^2$, at each dimension 
there are only a few of them.
At the dimension-four level, there is a single operator generated, i.e.,
the photon-mass operator:
\begin{equation}
\mira^2 A_\mu A^\mu\,.
\label{}
\end{equation}
At the dimension-six level, there are more operators, 
but only one, $\PP$, enters our ADM computation 
because $\Q_1$ and $\Q_3$ mix into it at one-loop.
It reads
\begin{equation}
\PP = \frac{1}{e}\mira^2 \sum_a \BPsi_a \gamma_\mu \Psi^a A^\mu\,.
\label{eq:poperator}
\end{equation}

\subsection{Renormalizing Green's functions\label{sec:greensfunctions}}

In this subsection, we highlight the relevant 
aspects in the computation of the renormalization constants $\Z_i^{~j}$, 
from which we extracted the ADM presented above, 
via the renormalization of amputated one-particle irreducible 
Green's functions.

\begin{table}[t]
\makebox[\textwidth][c]{
\begin{tabular}{lclc}
		&{\bf Green's function}	& \multicolumn{1}{c}{{\bf Depends on}} &	{\bf Constant(s) extracted} \\\hline\\[-0.8em]
{\bf One-loop}	& $A^\mu A^\nu$  	& $\dcZ{1}{\cO\N_2}$
					& $\dcZ{1}{\cO\N_2}$\\
		& $\BPsi\Psi A^\mu$	& $\dcZ{1}{\cO\N_1}$, $\dcZ{1}{\cO\PP}$, $\dcZ{1}{\cO\N_2}$
					& $\dcZ{1}{\cO\N_1}$, $\dcZ{1}{\cO\PP}$\\
		& $\BPsi\Psi\BPsi\Psi$	& $\dcZ{1}{\cO\cO'}$, $\dcZ{1}{\cO\N_1}$
					& $\dcZ{1}{\cO\cO'}$\\[2em]
{\bf Two-loop}	& $A^\mu A^\nu$  	& $\dcZ{2}{\Q\N_2}$, $\dcZ{1}{\Q\N}$, $\dcZ{1}{\Q,\PP}$
					& $\dcZ{2}{\Q\N_2}$\\
		& $\BPsi\Psi A^\mu$	& $\dcZ{2}{\Q\N}$, $\dcZ{1}{\Q\cO}$,  $\dcZ{1}{\Q\N}$,   $\dcZ{1}{\Q\PP}$
					& $\dcZ{2}{\Q\N_1}$\\
		& $\BPsi\Psi\BPsi\Psi$	& $\dcZ{2}{\Q\Q'}$, $\dcZ{2}{\Q\N_1}$, $\dcZ{1}{\Q\cO}$, $\dcZ{1}{\Q\N}$, $\dcZ{1}{\Q\PP}$
					& $\dcZ{2}{\Q\Q'}$\\[0.2em]\hline
\end{tabular}}
\caption{A summary of the Green's functions we consider.
The loop order ($L$-loop) refers to the $\alpha^L$ contribution to the corresponding
Green's function (second column). 
The third column contains the mixing renormalization
constants that the given Green's function depends on.
The last column contains the ones we extract in each case.
\label{tab:greensfunctions}
}
\end{table}

For each Green's function we need to specify the operator we insert
and the elementary fields on the external legs. 
In our case, the external legs are either four elementary
fermions, or two fermions and a photon, or two photons.
At the tree-level, a Wick contraction with the 
elementary fields defines a vertex structure for each operator.
We denote the $\overline{\Psi}\Psi\overline{\Psi}\Psi$ structures with $S$, 
the $\overline{\Psi}\Psi A^\mu$ ones with $\tilde{S}$, 
and the $A^\mu A^\nu$ one with $\hat{S}$. 
An additional subscript indicates the operator associated to 
a given structure. 
The representation in terms of Feynman diagrams is
 \begin{center}
\parbox{40mm}{\includegraphics{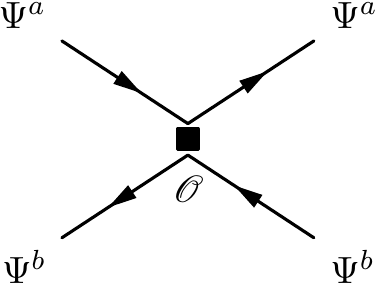}}$ = i  C^\cO S^{}_\cO$
\qquad\qquad
\parbox{40mm}{\includegraphics{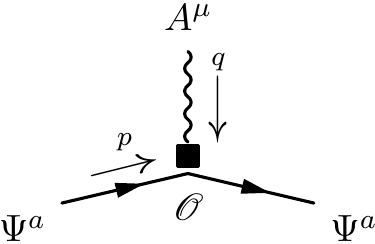}}$      = i  C^\cO \tilde S^{}_\cO$ \\[2em]
\parbox{40mm}{\includegraphics{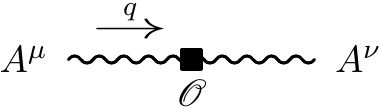}}$           = i  C^\cO \hat S^{}_\cO$ 
\end{center}
We collect all structures that enter the computation in 
appendix~\ref{app:feynmanrules}. 

In what follows, we refer to
\begin{equation}\label{eq:notation}
\BP{\cO}{L}{S}
\end{equation}
as a sum over a specific subset of Feynman diagrams:
{\it i)}  All these diagrams have a single insertion of the operator $\cO$.
{\it ii)} They are dressed with interactions such that they contribute
at $\order{\alpha^L}$.
In particular, we include all counterterm diagrams
proportional to field and charge renormalization constants,
but we {\it do not include} diagrams that contain mixing constants.
We keep those separate to demonstrate how we extract them.
{\it iii)} The subscript $S$ indicates that out of this sum of diagrams we only take
the part proportional to the structure $S$. 
In short, the notation of eq.~\eqref{eq:notation} denotes
{\it the $L$-loop insertion of $\cO$ projected on $S$}, including 
contributions from field and charge renormalization constants. 
\begin{figure}[]
\begin{center}
\begin{multline*}
\BP{\N_1}{2}{\tilde S} \equiv
\left[\parbox{10em}{\includegraphics{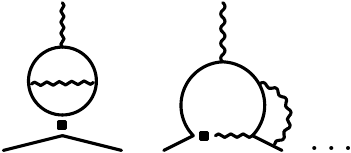}}\right]_{\tilde S}+
\left[\parbox{5.5em}{\includegraphics{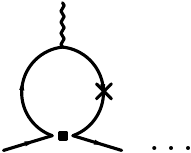}}\right]_{\tilde S}+
2 Z_\Psi^{(1)}		 \left[\parbox{5.5em}{\includegraphics{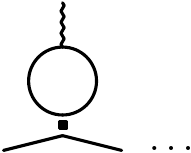}}\right]_{\tilde S}+\\+
(Z_\Psi^{(1)}+Z_A^{(1)}) \left[\parbox{5.5em}{\includegraphics{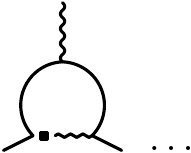}}\right]_{\tilde S}+
(Z_\Psi^{(2)}+Z_A^{(2)}+Z_\Psi^{(1)} Z_A^{(1)}) \left[\parbox{3.5em}{\includegraphics{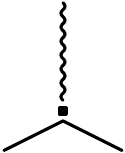}}\right]_{\tilde S}
\end{multline*}
\end{center}
\caption{An illustration of the non-trivial case of $\BP{\N_1}{2}{\tilde S}$, with
$\tilde S$ any of the structures in eq.~\eqref{eq:tildeStr}.
The squares denote operator insertions and the crosses counterterms.
The first parenthesis collects two-loop insertions of the operator $\N_1$,
which is a linear combination of $\N_1^\gamma$ and $\Q_1$.
The second collects the one-loop insertions with counterterms  on the
propagators and the QED vertices.
The third and fourth are one-loop insertions multiplied with the field and charge
renormalization of the fields and charges composing the $\N_1$, see
eq.~\eqref{eq:N1bare}.
The fifth are the tree-level insertions multiplied with the two-loop field and charge
renormalization constant from the $\N_1^\gamma$.
}
\label{fig:N1S2}
\end{figure}

As an illustration of the notation we show in figure~\ref{fig:N1S2} a small
subset of the Feynman diagrams for the non-trivial case of $\BP{\N_1}{2}{\tilde
S}$, with $\tilde S$ any of the structures in eq.~\eqref{eq:tildeStr}.
Notice that since $\N_1$ is a linear combination of terms with different fields,
see eq.~\eqref{eq:defN1}, its field and charge renormalizations depend on the part we
insert, namely
\begin{equation}
(\N_1)_0 = Z_\alpha^{-1/2}Z_A^{1/2}Z_\Psi 
\frac{1}{e} \partial^\nu F_{\mu\nu}(\BPsi_a\gamma^\mu \Psi^a) 
+ (Z_\Psi)^2 (\BPsi_a\gamma^\mu \Psi^a)^2\,.
\label{eq:N1bare}
\end{equation}

Next we derive the conditions on the Green's functions 
that determine the mixing constants.
For transparency we frame the constant(s) that  we extract
from a given condition.
In table~\ref{tab:greensfunctions} we summarize which Green's functions
we consider, on which mixing renormalization constants they depend, 
and which one we extract in each case.
For brevity we use the following shorthand notation:
\begin{align*}
&\Q,\,\Q'=\Q_1,\,\Q_3\,,&
&\E=\E_n\,,&
&\N=\N_1,\,\N_2\,,&
&\cO,\,\cO'=\Q_1,\,\Q_3,\,\E_n\,.&
\label{}
\end{align*}
We collect the results for the renormalization constants in
appendix~\ref{app:renormalizationconstants}.

\subsubsection*{$\boldsymbol{A^\mu A^\nu} $ at one-loop}
At one-loop there is no insertion of any four-fermion operator
that contributes to the Green's function with only two external photons.
Thus
\begin{equation}
\bdcZ{1}{\cO \N_2}=0\,.
\label{eq:dZ1XN2}
\end{equation}

\subsubsection*{$\boldsymbol{\overline{\Psi} \Psi A^\mu} $ at one-loop}
Contrarily, one-loop insertions of four-fermion operator 
contribute to the $\overline{\Psi} \Psi A^\mu $ Green's function.
By expanding the diagram in the basis of $\tilde{S}$ structures, we
determine the mixing into operators
with a tree-level projection onto $\BPsi\Psi A^\mu$, 
namely $\N_1$ and $\PP$.
For the physical operators the conditions are
\begin{align}
\BP{\Q}{1}{\tilde{S}_{\N_1}} +
\bdcZ{1}{\Q\N_1} \BP{\N_1}{0}{\tilde{S}_{\N_1}} +
\dcZ{1}{\Q\N_2} \BP{\N_2}{0}{\tilde{S}_{\N_1}}& = \order{\epsilon^0}\,,\\
 \BP{\Q}{1}{\tilde S_\PP} +
\bdcZ{1}{\Q\PP} \BP{\PP}{0}{\tilde S_\PP}& =\order{\epsilon^0} \,.
\label{}
\end{align}
In the first line we use that  
$\dcZ{1}{\Q\N_2} = 0$, as extracted from the $A^\mu A^\nu$ Green's function.
Similarly, we determine the mixing of $\E_n$ into $\N_1$\begin{align}
\BP{\E_n}{1}{\tilde{S}_{\N_1}} +
\bdcZ{1}{\E_n\N_1} \BP{\N_1}{0}{\tilde{S}_{\N_1}} +
\dcZ{1}{\E_n\N_2}\BP{\N_2}{0}{\tilde{S}_{\N_1}}& = \order{\epsilon}\,.
\end{align}
Notice that in this case the mixing constants subtract finite terms, 
as required by the flavor scheme we adopt.

\subsubsection*{$\boldsymbol{ \BPsi\Psi\BPsi\Psi}$ at one-loop}
Next, we compute the one-loop insertions in the $\BPsi\Psi\BPsi\Psi$ Green's
function.
Firstly, we insert physical operators, i.e., $\Q$,
\begin{align}
 \BP{\Q}{1}{S_\cO} +
\bdcZ{1}{\Q \cO} \BP{\cO}{0}{S_\cO} +
\dcZ{1}{\Q \N_1} \BP{\N_1}{0}{S_\cO} = & \order{\epsilon^0}\,,
\intertext{with the only non-vanishing $\BP{\N_1}{0}{S_\cO}$
being the one for $\cO=\Q_1$.
We see that extracting $\dcZ{1}{\Q\Q_1}$ assumes
knowledge of $\dcZ{1}{\Q\N_1}$, which we have previously 
determined via the $\BPsi\Psi A^\mu$ Green's function.
Next, we insert evanescent operators.
Again, the only difference here is that their mixing constants 
into physical operators subtract finite pieces
}
\BP{\E_n^{}}{1}{S_\Q} +
\bdcZ{1}{\E_n^{} \Q} \BP{\Q}{0}{S_\Q} +
\dcZ{1}{\E_n^{} \N_1} \BP{\N_1}{0}{S_\Q}= &  \order{\epsilon} \,,\\
\BP{\E_n^{}}{1}{S_{\E}} +
\bdcZ{1}{\E_n \E} \BP{\E}{0}{S_{\E}} = &  \order{\epsilon^0} \,.
\end{align}

This completes the computation of all one-loop constants required to
determine the mixing of physical operators at the two-loop level.
Next, we renormalize the same Green's functions at the two-loop level.

\subsubsection*{$\boldsymbol{A^\mu A^\nu}$ at two-loop}
At the two-loop order $\Q_1$ and $\Q_3$ insertions do contribute to the
$A^\mu A^\nu$ Green's function.
They can thus mix into the operator $\N_2$.
Even though  $\N_2$ itself does not have a tree-level projection on
physical operators, we need this mixing to extract the
two-loop mixing of $\Q_1$ and $\Q_3$ into $\N_1$ in the next step.
The projection onto the $\hat{S}$ structure results in the condition
\begin{equation}
\begin{split}
                      \BP{\Q}{2}{\hat S_{\N_2}} +
 \bdcZ{2}{\Q\N_2} \BP{\N_2}{0}{   \hat S_{\N_2}}+
 \sum_\N \dcZ{1}{\Q\N}   \BP{\N }{1}{ \hat S_{\N_2}} +
 \dcZ{1}{\Q\PP}    \BP{\PP}{1}{  \hat S_{\N_2}}
 = \order{\epsilon^0}\,.
\end{split}
\label{}
\end{equation}

\subsubsection*{$\boldsymbol{\overline{\Psi} \Psi A^\mu}$ at two-loop}
Next we renormalize the $\BPsi\Psi A^\mu$ Green's function at the two-loop level.
We only need the two-loop mixing of physical operators into $\N_1$,
because only $\N_1$ has a tree-level projection onto $\Q_1$.
To unambiguously determine the projection on the structure
$\tilde S_{\N_1}$, we have to fix a basis of linear independent structures,
which correspond to linearly independent operators.
At this loop order, we find that apart from $\N_1$ we also need to include
the operators $\N_3$ and $\N_4$ to project all generated structures.
This projection is the only point in which  these operators enter our
computation.
The finiteness of the two-loop $\BPsi\Psi A^\mu$ Green's function determines 
the two-loop mixing of physical operators into $\N_1$ via
\footnote{
Note that 
$\BP{\N_1}{1}{\tilde S_{\N_1}} =
\BP{\N_1^\gamma}{1}{\tilde S_{\N_1}} +
\BP{\Q_1}{1}{\tilde S_{\N_1}}\,,
$
as $\N_1$ has two Feynman rules.}
\begin{multline}
 \BP{\Q}{2}{                      		\tilde S_{\N_1}}+
\bdcZ{2}{\Q \N_1} \BP{\N_1}{0}{                 \tilde S_{\N_1}}+
\dcZ{2}{\Q \N_2} \BP{\N_2}{0}{                  \tilde S_{\N_1}}+\\+
\sum_{\cO}\dcZ{1}{\Q \cO} \BP{\cO}{1}{          \tilde S_{\N_1}}+
\sum_{\N}\dcZ{1}{\Q \N} \BP{\N}{1}{ 	        \tilde S_{\N_1}}+
\dcZ{1}{\Q \PP} \BP{\PP}{1}{        		\tilde S_{\N_1}}
= \order{\epsilon^0}\,.
\label{eq:ffarenormalization}
\end{multline}

\subsubsection*{$\boldsymbol{\BPsi\Psi\BPsi\Psi}$ at two-loop}
Finally, we have collected all results necessary to renormalize the
two-loop $\BPsi\Psi\BPsi\Psi$ Green's function.
The renormalization conditions for the mixing in the physical sector read
\begin{align}
&\BP{\Q}{2}{S_{\Q_1}} +
\bdcZ{2}{\Q \Q_1}  \BP{\Q_1 }{0}{S_{\Q_1}}+
\dcZ{2}{\Q \N_1} \BP{\N_1 }{0}{S_{\Q_1}} +
\sum_\cO\dcZ{1}{\Q \cO} \BP{\cO }{1}{S_{\Q_1}} +\label{eq:ffffQ1}\\
&\hspace{11.25em}+\sum_\N\dcZ{1}{\Q \N} \BP{\N }{1}{S_{\Q_1}} +
\dcZ{1}{\Q \PP} \BP{\PP}{1}{S_{\Q_1}}
= \order{\epsilon^0}\,,\nonumber\\
&\BP{\Q}{2}{S_{\Q_3}} +
\bdcZ{2}{\Q \Q_3}  \BP{\Q_3 }{0}{S_{\Q_3}}+
\sum_\cO\dcZ{1}{\Q \cO} \BP{\cO }{1}{S_{\Q_3}}+
\sum_\N\dcZ{1}{\Q \N} \BP{\N }{1}{S_{\Q_3}}
+\label{eq:ffffQ3}\\
&\hspace*{21em}+\dcZ{1}{\Q \PP} \BP{\PP}{1}{S_{\Q_3}}
= \order{\epsilon^0}\,.\nonumber
\end{align}
We see here explicitly that, because $\N_1$ has a tree-level projection onto $\Q_1$,
we need $\dcZ{2}{\Q \N_1}$ to determine $\dcZ{2}{\Q \Q_1}$.

\subsection{Evaluation of Feynman diagrams\label{sec:Loops}}
Already at the two-loop level the number of Feynman 
diagrams entering the Green's functions is quite large.
The present computation is thus performed in an  automated setup.
Firstly, the program {\tt QGRAF} \cite{Nogueira:1991ex} generates
all diagrams creating a symbolic output for each diagram.
This output is converted to the algebraic structure of a loop
diagram and subsequently computed using self-written routines in {\tt FORM}
\cite{Kuipers:2012rf}.
The methods for the computation and extraction of the UV poles of two-loop
diagrams are not novel and also widely used throughout the literature.
Here, we shall only sketch the steps and mention parts specific to our
computation.

One major simplification of the computation comes from the fact that we can always
expand the integrand in powers of external momenta over loop-momenta and drop terms
beyond the order we are interested in.
For instance, for the $\BPsi\Psi\BPsi\Psi$ Green's function all 
external momenta can be directly set to zero, 
while for the $\BPsi\Psi A^\mu$ one we need to keep the  external momenta
up to second order to obtain the mixing into $\N_1$ (see $\tilde S_{\N_1}$ in
eq.~\eqref{eq:tildeStr}).

After the expansion, all propagators are massless so the resulting loop-integrals
vanish in dimensional regularization.  
To regularize the IR poles and perform the expansion in external
momenta we implement the ``Infrared Rearrangement'' (IRA) procedure introduced in
refs.~\cite{Misiak:1994zw,Chetyrkin:1997fm}.
In IRA, an ---in our case massless--- propagator is replaced using the identity
\begin{equation}
\frac{1}{(p+q)^2}  =  \frac{1}{p^2-\mira^2}  - \frac{q^2+2 p\cdot q + \mira^2}{p^2-\mira^2}\frac{1}{(p+q)^2}\,,
\label{}
\end{equation}
where $p$ is the loop momentum, $q$ is a linear combination of 
external momenta, and $\mira$ is an artificial, unphysical mass.
We see that the first term in the decomposition contains the scale $\mira$ and
carries no dependence on external momenta in its denominator.
In the second term, the original propagator reappears, but thanks to the additional
factor the overall degree of divergence of the diagram is reduced by one.
When we apply the decomposition multiple times, we obtain
a sum of terms with only loop-momenta and $\mira$ in the denominators plus
a term proportional to $\frac{1}{(p+q)^2}$.
This last term, however, can be made to have 
an arbitrary small degree of divergence. Therefore, in a given diagram we can always perform the decomposition as many
times as necessary until terms proportional to  $\frac{1}{(p+q)^2}$ are finite
and can thus be dropped if we are interested in UV poles.

When applying IRA on photon propagators, the 
resulting coefficients of the poles are not gauge-invariant, because
we drop the finite terms in the expansion of propagators.
This is why some gauge-variant operators/counterterms enter in
intermediate stages of the computation, for instance the operator $\PP$.
Such operators are always proportional to $\mira^2$ and so only a 
small number of them enters at each dimension.
For more details on the prescription 
we refer to the original work~\cite{Chetyrkin:1997fm}.

The IRA procedure results in integrals with denominators 
that {\it i)} are independent from external momenta,
and {\it ii)} contain the artificial mass $\mira$.
We can always reduce these integrals to scalar ``vacuum'' diagrams
by contracting them with metric tensors and solving the resulting system of linear
equations, e.g., see  ref.~\cite{Chetyrkin:1997fm}.
This tensor reduction reduces all integrals to one- and two-loop
scalar integrals of the form
\begin{equation}
\int\!     \frac{d^d p}{(p^2-m_1^2)^{n_1}}
\qquad\text{and}\qquad
\int\!\!\!\int\! \frac{d^d p_1 d^d p_2}{(p_1^2-m_1^2)^{n_1}(p_2^2-m_2^2)^{n_2}(p_1-p_2)^{2n_3}}\,,
\label{}
\end{equation}
with the integers $n_1$, $n_2$, $n_3\ge 1$, and $m_1\neq 0$.
The one-loop integral can be directly evaluated, whereas all two-loop
integrals can be reduced to a few  master integrals 
using the recursion relation
in ref.~\cite{Bobeth:1999mk}. In fact, in our case $m_1=m_2=\mira$ and the use of recursion relations is not required.

In the evaluation of the Feynman diagrams, we use 
the Clifford algebra in $d$ dimensions for 
{\it i)} the evaluation of traces with gamma matrices
when the diagram in question has closed fermion loops, 
and {\it ii)} the reduction of the Dirac structures to 
the operator structures $S$ or $\tilde{S}$ listed in 
appendix~\ref{app:feynmanrules}.

\subsection{Anomalous dimensions at the fixed point\label{sec:evanescent}}
By substituting the value of the coupling at the fixed point, eq.~\eqref{eq:alphaStar},
in the result of eq.~\eqref{eq:anodimalphaff}, 
we obtain the ADM at the fixed point as an expansion  in $\epsilon$
\begin{equation}
\gamma^* = \gamma_1^* \,\epsilon + \gamma_2^* \,\epsilon^2 + \order{\epsilon^3}~,
\end{equation}
where
\begin{align}
(\gamma_1^*)_{nm} &= \frac{3}{2 \Nf}\times
\left\{
\begin{array}[]{cl}
8 \delta_{n3} + n(n-1)(n-5)(n-6)&\text{for}\quad m=n-2\\
\frac 43 (2 \Nf +1)\delta_{n1}-2(n-1)(n-3)	&\text{for}\quad m=n\\
1 		&\text{for}\quad m=n+2~\\
0               & \text{otherwise}~,
\end{array}
\right.\label{}\\[1em]
(\gamma_2^*)_{nm} &=
\left\{
\begin{array}[]{ll}
	-\frac{1}{24 \Nf^2}
	\begin{bmatrix}
	2383 + 224 \Nf 		& 375 + 18 \Nf\\
	-1212 - 2568 \Nf	& -1485 - 360 \Nf
	\end{bmatrix}&\\
	+\frac{3}{8\Nf^2}
	a_5
	\begin{bmatrix}
	-3 & 0\\
	4\Nf+4 &3
	\end{bmatrix}
	+\frac{3}{8\Nf^2}
	b_5
	\begin{bmatrix}
	0 & -3\\
	132 & -4\Nf
	\end{bmatrix}
	&\text{for}~n,m=1,3~,\\[2em]
	\frac{24}{\Nf} (-1)^{\frac{n(n-1)}{2}} (n-2)(n-5)!\\
	\quad + \frac{3}{2 \Nf}  \left( -n(n-1)(n-5)(n-6) a_{n-2} \right. \\
	\qquad +\left( \frac{4}{3}(2\Nf+1)+ 2 (n-1)(n-3)\right) a_{n}\\
	\qquad\quad \left. -  a_{n+2} + 44 b_n \right)
	&\text{for}~m=1,\,n\ge5~,\\[1em]
	- \frac{60}{\Nf} \delta_{n5}\\
	\quad+ \frac{3}{2 \Nf}\left( - n(n-1)(n-5)(n-6)  b_{n-2} \right.\\
	\qquad\left. + 2(n-1)(n-3) b_{n}- b_{n+2}+ a_n \right)
	&\text{for}~m=3,\,n\ge5~,\\[1em]
\hspace*{7em}\text{not required} &\text{otherwise\,.}
\end{array}
\right. 
\end{align}
Note that the physical--physical block is not invariant 
at order $\epsilon^2$, because there are non-zero entries $(\gamma^*)_{n1}$ 
and $(\gamma^*)_{n3}$ for all $n\geq 5$. 

We are interested in finding the first two eigenvalues of $\gamma^*$
up to order $\epsilon^2$. 
They determine the scaling dimensions
of the corresponding eigenoperators at the IR fixed point.
We denote these scaling dimensions by
\begin{align}\label{eq:D1D2i}
(\Delta_{\IR})_i & = \Delta_{\UV}(\epsilon) + \epsilon (\Delta_1)_i + \epsilon^2 (\Delta_2)_i  + \order{\epsilon^3}~,
\end{align}
with $i = 1,2$ and $\Delta_{\UV}(\epsilon)=6-4\epsilon$.
To compute the first two eigenvalues we have truncated the problem to include a large 
but finite number of evanescent operators. Taking a sufficiently large truncation,
 the scheme/basis-dependence 
 of the approximated result can be made negligible
 at the level of precision we are interested in
(for details see ref.~\cite{DiPietroStamou1}).
In table~\ref{tab:delta12}, we list the values of
$(\Delta_1)_i$ and  $(\Delta_2)_i$ for $\Nf=1,\dots,10$
after we included enough evanescent operators such that the three
significant digits listed remain unchanged.
The table is the main result of this section. 
In section \ref{sec:d3extrapolation}, we will use these 
results as a starting point 
to extrapolate the scaling dimensions to $d=3$. 

\begin{table}
\makebox[\textwidth][c]{
\begin{tabular}{lrrrrrrrrrr}
\multicolumn{1}{r}{$\boldsymbol{\Nf}$} 		& $1$ & $2$ & $3$ & $4$ & $5$ & $6$ & $7$ & $8$ & $9$ & $10$ \\
\hline\hline
$\boldsymbol{(\Delta_{1})_1}$   & $-7.39$ & $-3.07$ & $-1.72$ & $-1.10$ & $-0.766$ & $-0.562$ & $-0.429$ & $-0.337$ & $-0.272$ & $-0.224$\\
$\boldsymbol{(\Delta_{2})_1}$   & $46.1$ & $14.1$ & $7.43$ & $4.84$ & $3.51$ & $2.73$ & $2.21$ & $1.86$ & $1.59$ & $1.39$\\[0.5em]
$\boldsymbol{(\Delta_{1})_2}$   & $13.4$ & $8.07$ & $6.39$ & $5.60$ & $5.17$ & $4.90$ & $4.71$ & $4.59$ & $4.49$ & $4.42$\\
$\boldsymbol{(\Delta_{2})_2}$   & $-84.0$ & $-23.5$ & $-11.6$ & $-7.12$ & $-4.94$ & $-3.70$ & $-2.91$ & $-2.37$ & $-1.99$ & $-1.70$\\\hline
\end{tabular}}
\caption{
The values of the one-loop $(\Delta_1)_i$ and the two-loop $(\Delta_2)_i$ coefficients
defined in eq.~\eqref{eq:D1D2i} for $\Nf=1,\dots,10$.
Only three significant digits are being displayed.
\label{tab:delta12}
}
\end{table}

\section{\texorpdfstring{Bilinear operators in $\boldsymbol{d=4-2\epsilon}$\label{sec:bilinearsind4m2ep}}{%
	 Bilinear operators in d=4-2ep}}

In this section we consider operators that are
bilinear in the fermionic fields. The most generic 
bilinear operators without derivatives are 
\begin{align}
\BPsi_a \Gamma^n_{\mu_1\dots\mu_n}\Psi^b\,,&&
\BPsi_a \Gamma^n_{\mu_1\dots\mu_n}\gamma_5\Psi^b\,,
\label{eq:genericbilinears}
\end{align}
with $n\ge0$.  
$\gamma_5$ can be consistently continued to $d = 4-2\epsilon$ using the 
't\,Hooft--Veltman prescription \cite{tHooft:1972tcz, Breitenlohner:1977hr}. 
The indices $a,b = 1, \dots, \Nf$ 
are indices in the fundamental of the diagonal ``vector'' $SU(\Nf)$
subgroup of the $SU(\Nf)_L \times SU(\Nf)_R$ symmetry of the theory in $d=4$. 
In $d = 4-2\epsilon$, the conservation of the 
nonsinglet axial currents is violated by evanescent operators 
\cite{Collins:1984xc}, and thus only the diagonal 
$SU(\Nf)$ is a symmetry. 
On the other hand, the CFT in $d= 3$ is expected to enjoy the full 
$SU(\Nf)_L \times SU(\Nf)_R$ symmetry, which is actually 
enhanced to $SU(2 \Nf)\times U(1)$. 
Therefore, in continuing the operators of eq.~\eqref{eq:genericbilinears} 
to $d = 3$, we find that the ones with $\gamma_5$ are in the 
same multiplets of the flavor symmetry as those without.
So even though their scaling dimensions can differ as 
a function of $\epsilon$, the enhanced symmetry entails 
that they should agree when  $\epsilon = \frac{1}{2}$. Since the operators with $\gamma_5$ do not provide new 
information about the $3d$ CFT, and the 't\,Hooft--Veltman 
prescription makes computations technically more involved, 
we restrict our discussion here to operators without $\gamma_5$. 
As a future direction, it would be interesting to test
this prediction of the enhanced symmetry by comparing 
the scaling dimensions of operators with $\gamma_5$ after 
extrapolating to $d=3$ at sufficiently high order. 
We also restrict the discussion to operators with $n \leq 3$, 
because the others are evanescent in $d=3$. 

The anomalous dimension of bilinear operators without $\gamma_5$ 
has been computed for a generic gauge group
at three-loop accuracy in ref.~\cite{Gracey:2000am}.
For our $U(1)$ gauge theory we substitute $C_A= 0$ and
$C_F = T_F = 1$. 
Moreover, there is a difference in the normalization convention 
for the anomalous dimension, so that $\gamma^{\text{here}} = 2 
\gamma^{\text{there}}$. 
Under $SU(\Nf)$ each operator decomposes into a 
singlet and an adjoint component,
\begin{align}
   \BilL{n}{sing}{\mu_1\dots\mu_n} 		& =\BPsi_a\Gamma^n_{\mu_1\dots\mu_n}\Psi^a ~, \\
(\,\BilL{n}{adj} {\mu_1\dots\mu_n}\,)_a^{~b}	& =\BPsi_a\Gamma^n_{\mu_1\dots\mu_n}\Psi^b - \frac{1}{\Nf} \delta_a^b \,\BPsi_c \Gamma^n_{\mu_1\dots\mu_n} \Psi^c  ~,
\end{align}
respectively. 
A priori, the two components can have different anomalous dimensions.
The difference between the 
singlet and the adjoint originates from diagrams 
in which the operator is inserted in a closed fermion loop. 
When the operator has an even number of gamma matrices, 
the closed loop gives a trace with an odd total number of 
gamma matrices, which vanishes.
So for even $n$ there is no difference between the singlet and the
adjoint, i.e., they have the same anomalous dimension.

Below we collect the results for $n \leq 3$.
\begin{description}
\item[Scalar:] 
\begin{multline}
 \Delta(\Bil{0}{sing}{}) = 
 \Delta(\Bil{0}{adj}{}) = 
  3 - 2 \epsilon 
  -\frac{9}{2 \Nf} \epsilon 
  + \frac{60 \Nf+ 135}{16 \Nf^2} \epsilon^2\\ 
  + \frac{140 \Nf^2-81 \Nf (16 \zeta (3)-5)-3078}{32 \Nf^3} \epsilon^3 + \order{\epsilon^4}~.
\end{multline}
\item[Vector:] For $n=1$ both operators are conserved currents, 
so they do cannot have an anomalous dimension, i.e., 
\begin{align}
 \Delta(\Bil{1}{sing}{\mu}) = 
 \Delta(\Bil{1}{adj}{\mu}) = 
 3 - 2 \epsilon ~.
\end{align}
\item[Two-form:]
\begin{multline}
 \Delta(\Bil{2}{sing}{\mu\nu}) = 
 \Delta(\Bil{2}{adj}{\mu\nu}) = 
 3 - 2 \epsilon 
 +\frac{3}{2 \Nf} \epsilon
  - \frac{52 \Nf+ 225}{16 \Nf^2} \epsilon^2 \\
  - \frac{36 \Nf^2 - 3 \Nf (144 \zeta (3) + 287) + 1728 \zeta (3) - 3078}{32 \Nf^3} \epsilon^3 + \order{\epsilon^4}~.
\end{multline}
\item[Three-form:] 
\begin{align}
 \Delta(\Bil{3}{sing}{\mu\nu\rho}) &=  3 - 2 \epsilon  + \frac{15}{2 \Nf} \epsilon^2 + \frac{26 \Nf - 369}{8 \Nf^2} \epsilon^3 + \order{\epsilon^4}~, \\
 \Delta(\Bil{3}{adj}{\mu\nu\rho})  &=  3 - 2 \epsilon - \frac{6}{\Nf} \epsilon^2 + \frac{\Nf+ 45}{\Nf^2} \epsilon^3 + \order{\epsilon^4}~.
\end{align}
In $d = 4$ these three-form operators are Hodge-dual to axial currents. 
Actually, the fact that they do not get 
an anomalous dimension at one-loop, as seen from the equations above, is  related to 
this.
However, Hodge-duality cannot be defined in 
$d = 4 - 2\epsilon$ and the anomalous dimensions 
start to differ from those of the 
axial current at the two-loop level.
\end{description}
This exhausts the list of bilinears without $\gamma_5$
that flow to physical operators as $d\to3$.
In section \ref{sec:bilinearsd3} we discuss which operators of the 
CFT in $d = 3$ are continued by the operators above, and extrapolate 
the above results to obtain estimates for their scaling dimensions. 

\section{\texorpdfstring{Extrapolation to $\boldsymbol{d=3}$\label{sec:d3extrapolation}}{%
	 Extrapolation to d=3}}

\subsection{Pad\'e approximants\label{sec:pade}}

A computation of a certain order in $\epsilon$ provides  an 
approximation to the observable, e.g. the scaling 
dimension $\Delta$, in terms of a polynomial
\begin{equation}
\Delta = \Delta_{\UV}(\epsilon) + \sum_{i = 1}^k\Delta_i\epsilon^i 
+ \order{\epsilon^{k+1}}~.
\label{eq:fixedorder}
\end{equation}

Taking $\epsilon\to \frac{1}{2}$ in this polynomial  
gives the ``fixed order'' $d=3$ prediction of the $\epsilon$-expansion. 
Typically, the fixed-order results show
poor convergence as the order is increased.
A standard resummation technique adopted for these kind of 
extrapolations is to replace the polynomial with a Pad\'e approximant. 
The Pad\'e approximant of order ($k$,$l$) is defined as
\begin{equation}
\Delta^{\text{Pad\'e}}(k,l)\equiv
\frac{\sum_{i=0}^k c_i\epsilon^i}{1+\sum_{j=1}^l d_j\epsilon^j}\,.
\label{eq:pade1}
\end{equation}
The coefficients $c_i$ and $d_i$ are determined by 
matching the expansion of eq.~\eqref{eq:pade1} with
eq.~\eqref{eq:fixedorder}. $k+l$ must equal
the order at which we are computing.
Another condition comes from the fact 
that we are interested in the result for $\epsilon\to
\frac{1}{2}$.
In order for the $\epsilon$-expansion to smoothly interpolate 
from $\epsilon=0$ to $\epsilon=\frac{1}{2}$, an employable 
Pad\'e approximant should not have poles for
$\epsilon\in[0,\frac{1}{2}]$ for the values of $\Nf$ that
we consider.
In what follows, we show the predictions from a Pad\'e approximation
only if it does not contain any pole on the positive axis of
$\epsilon$ for any value of $\Nf=1,\dots,10$.

\subsection{\texorpdfstring{Four-fermion operators as ${d\to 3}$\label{sec:4fermiond3}}{%
	    Four-fermion operators as d->3}}

\begin{table}
\makebox[\textwidth][c]{
\begin{tabular}{llrrrrrrrrrr}
&\multicolumn{1}{r}{$\boldsymbol{\Nf}$} 		& $1$ & $2$ & $3$ & $4$ & $5$ & $6$ & $7$ & $8$ & $9$ & $10$ \\
\hline\hline
$\boldsymbol{(\Delta)_{1}}$
& {\bf LO}	          & $0.304$ & $2.47$ & $3.14$ & $3.45$ & $3.62$ & $3.72$ & $3.79$ & $3.83$ & $3.86$ & $3.89$\\
& {\bf NLO Pad\'e (1,1)}  & $4.12$ & $4.23$ & $4.27$ & $4.27$ & $4.26$ & $4.24$ & $4.23$ & $4.21$ & $4.20$ & $4.19$\\[0.5em]
$\boldsymbol{(\Delta)_{2}}$
&{\bf LO}		  & $10.7$ & $8.03$ & $7.19$ & $6.80$ & $6.58$ & $6.45$ & $6.36$ & $6.29$ & $6.25$ & $6.21$\\
&{\bf NLO Pad\'e (1,1)}   & $6.86$ & $6.52$ & $6.35$ & $6.25$ & $6.19$ & $6.15$ & $6.12$ & $6.10$ & $6.08$ & $6.07$\\\hline
\end{tabular}}
\caption{
LO and NLO Pad\'e (1,1) predictions for the scaling dimension of the two 
flavor-singlet four-fermion operators at $d=3$ for various values of $\Nf$.
Only three significant digits are being displayed.
\label{tab:delta12LOandNLO}
}
\end{table}

In $d=3$, the two four-fermion operators in the UV can be rewritten as
\begin{equation}
\Q_1 \underset{d \to 3}{\longrightarrow} (\overline{\psi}_i \gamma^{(3)}_\mu \psi^i)^2~, \quad 
\Q_3 \underset{d \to 3}{\longrightarrow} (\overline{\psi}_i \psi^i)^2\,,
\end{equation}
where $i=1,\dots,2\Nf$.
In this rewriting we see explicitly that these operators are singlets
of $SU(2 \Nf)$.
 
We now evaluate the scaling dimensions $(\Delta)_1$ and $(\Delta)_2$ of the two 
corresponding IR eigenoperators, at NLO.
For the NLO prediction we employ the Pad\'e approximation of order (1,1).
We list the values of the LO and NLO Pad\'e (1,1) predictions for the values of
$\Nf=1,\dots,10$ in table~\ref{tab:delta12LOandNLO}.

\begin{figure}[]
\begin{center}
\includegraphics{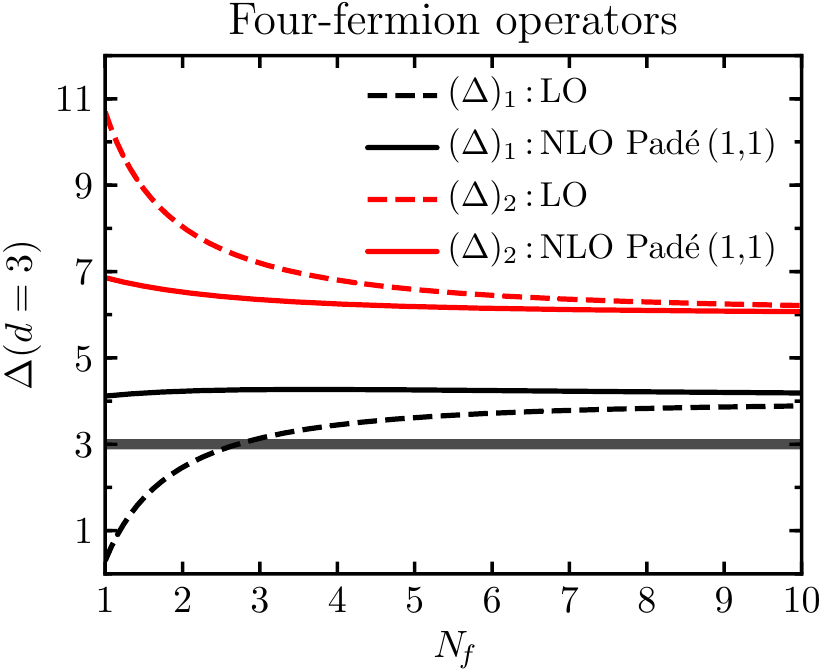}
\end{center}
\caption{Extrapolations of the scaling dimensions 
of the two flavor-singlet four-fermion operators to $d=3$, as a function of $\Nf$.
In black (lower two lines) $(\Delta)_1$ and in red (upper two lines) $(\Delta)_2$. 
Dashed lines are the LO estimate and 
solid lines the NLO Pad\'e (1,1).  
\label{fig:DeltaNf}
}
\end{figure}

We visualize the results in figure~\ref{fig:DeltaNf}.
The dashed lines are the result of the one-loop $\epsilon$-expansion computation.
Indeed, as discussed in ref.~\cite{DiPietro:2015taa}, the one-loop approximation
predicts that the lowest eigenvalue becomes relevant for $\Nf<3$.
The two-loop computation presented here changes this prediction.
The two solid lines represent the NLO Pad\'e (1,1) approximation to the 
two scaling dimensions.
We observe that for no value of $\Nf$ does the lowest eigenvalue 
reach marginality. We also see that the corrections to the 
LO result are significant, especially for small $\Nf$, i.e., $\Nf=1,2$.
This means that for such small values of $\Nf$, 
NLO accuracy is not sufficient to obtain a precise 
estimate for this scaling dimension.
Nevertheless, at face value, the result of the two-loop $\epsilon$-expansion 
suggests that  QED$_3$ is conformal in the IR for any value of $\Nf$. 

\begin{figure}[]
\begin{center}
\includegraphics{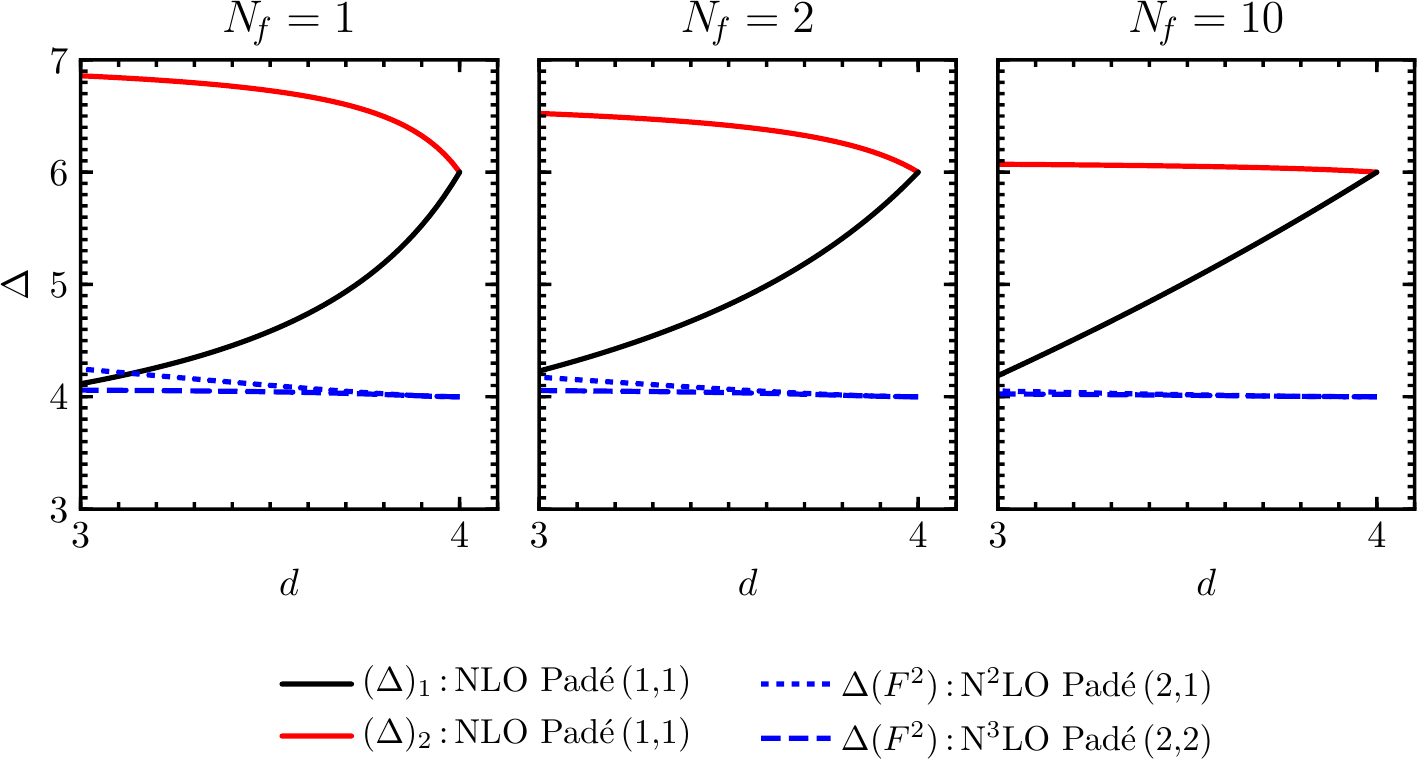}
\end{center}
\caption{$\epsilon$-expansion approximations to the scaling
dimensions of the two flavor-singlet four-fermion operators (black and red lines) 
and $F_{\mu\nu}^2$ (blue lines) as a function of the dimension $d$, i.e., for
$\epsilon\in[0,\frac{1}{2}]$.
The left, center, and right panel show the result for
the representative cases of $\Nf=1,2,$ and $10$, respectively.
We observe that the N$^3$LO Pad\'e (2,2) prediction of $\Delta(F^2)$ 
never crosses the  NLO Pad\'e (1,1) prediction of $(\Delta)_1$ 
in the extrapolation region.
\label{fig:F2diagonal}}
\end{figure}

Next, we comment on the relation of our result to the 
$1/\Nf$-expansion in $d=3$.
At large $\Nf$, the gauged $U(1)$ current, $\overline{\psi}_i \gamma^{(3)}_\mu \psi^i$, 
is set to zero by the EOM of the gauge field,
hence the operator $\Q_1$ is an EOM-vanishing 
operator. 
However, besides $\Q_3$, there still is another flavor-singlet scalar operator
of dimension $4$ for $\Nf = \infty$, namely $F_{\mu\nu}^2$.
$\Q_3$ and $F_{\mu\nu}^2$ mix at 
order $1/\Nf$ \cite{Chester:2016ref}.
Looking at the $\epsilon$-expansion result in figure~\ref{fig:DeltaNf}
we see that indeed only the lowest eigenvalue $(\Delta)_1$ (black lines) 
approaches $4$ for large $\Nf$.
The other scaling dimension (red lines) approaches $6$ as $\Nf\to \infty$, 
implying that the two eigenoperators cannot mix at large $\Nf$. 
This is consistent precisely 
because there is only one non-trivial singlet four-fermion operator 
at large $\Nf$.  
Its mixing with $F_{\mu\nu}^2$ cannot be captured within the 
$\epsilon$-expansion, because the UV dimension of $F_{\mu\nu}^2$ 
differs from that of a four-fermion operator in $d= 4- 2\epsilon$.
We can, however, test whether for any value of $\epsilon\in[0,\frac{1}{2}]$ 
the lowest eigenvalue $(\Delta)_1$, which starts off larger at $\epsilon = 0$, 
crosses the dimension of $F_{\mu\nu}^2$. 
Such a level-crossing would require to revisit the extrapolation to
$\epsilon = \frac{1}{2}$ and possibly affect the estimate. 
The scaling dimension of $F_{\mu\nu}^2$ in $\epsilon$-expansion is 
\begin{equation}
\Delta(F^2) = 4-2\epsilon + 
\alpha^2\frac{\partial}{\partial \alpha}\left.\left( \frac{1}{\alpha^2}\frac{d \alpha}{d\log\mu} \right)\right\vert_{\alpha=\alpha^*}\,,
\label{}
\end{equation}
with $\alpha^*$ given in eq.~\eqref{eq:alphaStar} up to 
$\order{\epsilon^4}$. 
At three- and four-loop order the only 
Pad\'e approximation without poles in the 
positive real axis of $\epsilon$ is the order (2,1) and (2,2), respectively.
In figure~\ref{fig:F2diagonal} we plot
$(\Delta)_{1,2}$ and $\Delta(F^2)$ as a function of $d$ for the representative cases of 
$\Nf=1,2,$ and $10$.
We observe that the only case in which $(\Delta)_1$ crosses 
$\Delta(F^2)$ before $d=3$ is when $\Nf=1$ and when we employ 
N$^2$LO Pad\'e (2,1) to predict $\Delta(F^2)$. 
The N$^3$LO Pad\'e (2,2) prediction for $\Nf=1$ does not
cross $(\Delta)_1$ and the same holds for larger values of $\Nf$.
Therefore, at least at this order, 
$F_{\mu\nu}^2$ should not play a significant role in obtaining the four-fermion scaling dimension.

\subsection{\texorpdfstring{Bilinears as ${d\to 3}$\label{sec:bilinearsd3}}{%
	    Bilinears as d->3}}

Next we consider bilinear operators in $d=3$. In the UV, restricting to the 
ones without derivatives, the possibilities are
\begin{description}
\item[Scalar:]
\begin{align}
 \cBil{0}{sing}{} 		& =  \overline{\psi}_i \psi^i\,, \\
(\,\cBil{0}{adj}{}\,)_i^{~j} 	& =  \overline{\psi}_i \psi^j - \frac{1}{2\Nf} \overline{\psi}_k \psi^k \delta_i^{~j}\,.
\end{align}
The subscript refers to the representation of $SU(2\Nf)$.
The singlet is parity-odd.
We can combine parity with an element of the Cartan of $SU(2 \Nf)$, 
in such a way that one component of the adjoint
scalar is parity-even. Since parity squares to the identity, this Cartan element
can only have $+1$ and $-1$ along the diagonal, which up to permutations we can take to be the first $\Nf$, and the second $\Nf$ diagonal entries, respectively.
With this choice, the parity-even bilinear is
$\sum_{a = 1}^{\Nf} (\overline{\psi}_a \psi^a - \overline{\psi}_{a+ \Nf} \psi^{a + \Nf})$.
This is the candidate to be the ``chiral condensate'' in QED$_3$ \cite{Vafa:1984xh}.
\item[Vector:]
\begin{align}
\cBilL{1}{sing}{\mu} 		& =  \overline{\psi}_i \gamma^{(3)}_\mu \psi^i\,, \\
(\,\cBilL{1}{adj}{\mu}\,)_i^{~j} & =  \overline{\psi}_i \gamma^{(3)}_\mu \psi^j - \frac{1}{2 \Nf} \overline{\psi}_k \gamma^{(3)}_\mu \psi^k \delta_i^{~j}\,.
\end{align}
The singlet is the current of the gauged $U(1)$. When the 
interaction is turned on, it recombines with the field strength and 
does not flow to any primary operator of the IR CFT.
The adjoint is the current that generates the $SU(2\Nf)$ global symmetry. 
Therefore, we expect it to remain conserved
along the RG and flow to a conserved current of dimension
$\Delta = 2$ in the IR.
\end{description}

We now identify 
which $d=4-2\epsilon$ bilinears from section~\ref{sec:bilinearsind4m2ep} approach
the $d=3$ bilinears above.
Substituting the decomposition of eqs.~\eqref{eq:Diracd3} and \eqref{eq:gammad3}, and 
also using $3d$ Hodge duality, we find that
\begin{align}
\Bil{3}{sing}{\mu\nu\rho}  		
	& \quad\overset{d\to 3}{\longrightarrow}\quad \cBil{0}{sing}{}\,,\\
\Bil{0}{adj}{},~\Bil{3}{adj}{\mu\nu\rho} 
	& \quad\overset{d\to 3}{\longrightarrow}\quad \cBil{0}{adj}{}\,,\\
\Bil{1}{sing}{\mu}  		
	& \quad\overset{d\to 3}{\longrightarrow}\quad \cBil{1}{sing}{\mu}\,, \\
\Bil{1}{adj}{\mu},~\Bil{2}{adj}{\mu\nu} 
	& \quad\overset{d\to 3}{\longrightarrow}\quad \cBil{1}{adj}{\mu}\,.
\label{}
\end{align}
We denote by $\Delta(\mathcal{B})$ the scaling dimension of the operator in the IR CFT in $d=3$ that a certain bilinear $\mathcal{B}$ flows to.
Therefore, we expect 
\begin{align}
\Delta(\Bil{3}{sing}{\mu\nu\rho})
	&\quad\overset{d\to3}{\longrightarrow}\quad \Delta(\cBil{0}{sing}{})\,,\\
\Delta(\Bil{0}{adj}{}),~\Delta(\Bil{3}{adj}{\mu\nu\rho}) 
	& \quad\overset{d\to3}{\longrightarrow}\quad \Delta(\cBil{0}{adj}{})\,,\\
\Delta(\Bil{2}{adj}{\mu\nu}) 
	& \quad\overset{d\to3}{\longrightarrow}\quad 2\,.
	\label{eq:DeltaEquals2}
\end{align}
The last equation provides a test of the $\epsilon$-expansion and the first two 
provide estimates of the observables $\Delta(\cBil{0}{sing}{})$ and 
$\Delta(\cBil{0}{adj}{})$. To this end, we employ the
viable Pad\'e approximants 
for $\Nf=1,\dots,10$.
In table~\ref{tab:bilinear} we list the $\epsilon$-expansion 
predictions at $d=3$ for $\Nf=1,\dots,10$.
For the cases in which the order (1,1) Pad\'e approximant is singular,
we list the fixed-order NLO prediction.

\begin{figure}[]
\begin{center}
\includegraphics{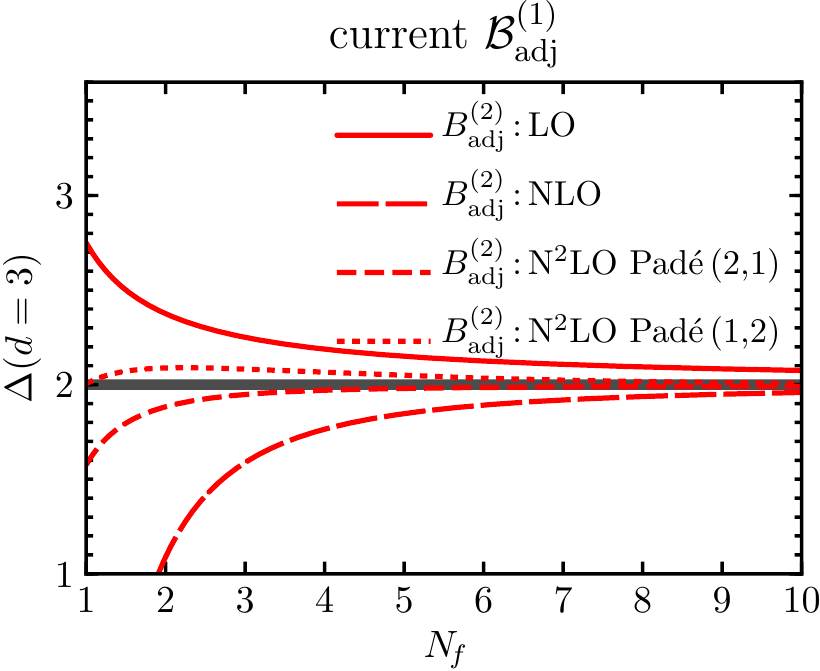}
\end{center}
\caption{%
$\epsilon$-expansion predictions for the scaling dimension of the bilinear
vector operator $\cBil{1}{adj}{\mu}$ at $d=3$.
The operator is associated to the conserved flavor-nonsinglet current of ~$SU(2\Nf)$, 
thus its scaling dimension is expected to equal $2$.
We observe that the N$^2$LO Pad\'e approximations are indeed 
close to this expectation even for small values of $\Nf$.
\label{fig:DeltaBilinearsNftest}
}
\end{figure}

\begin{table}

\makebox[\textwidth][c]{
\begin{tabular}{llrrrrrrrrrr}
& \multicolumn{1}{r}{$\boldsymbol{\Nf}$}& 		 $1$ & $2$ & $3$ & $4$ & $5$ & $6$ & $7$ & $8$ & $9$ & $10$ \\\hline\hline\\[-0.5em]
$\boldsymbol{\Delta(\cBil{0}{sing}{})}$ & $\boldsymbol{\Delta(\Bil{3}{sing}{\mu\nu\rho})}$ &&&&&&&&&&\\
&{\bf  LO}			& $2$ & $2$ & $2$ & $2$ & $2$ & $2$ & $2$ & $2$ & $2$ & $2$\\
&{\bf NLO Pad\'e  (1,1)}	& $2.65$ & $2.48$ & $2.38$ & $2.32$ & $2.27$ & $2.24$ & $2.21$ & $2.19$ & $2.17$ & $2.16$\\
&{\bf N$^2$LO Pad\'e (2,1)}	& $2.49$ & $2.40$ & $2.35$ & $2.30$ & $2.27$ & $2.24$ & $2.22$ & $2.20$ & $2.19$ & $2.17$\\
&{\bf N$^2$LO Pad\'e (1,2)}     & $2.51$ & $2.42$ & $2.35$ & $2.30$ & $2.27$ & $2.24$ & $2.22$ & $2.20$ & $2.18$ & $2.17$\\[1em]
$\boldsymbol{\Delta(\cBil{0}{adj}{})}$ & $\boldsymbol{\Delta(\Bil{0}{adj}{})}$ &&&&&&&&&&\\       
&{\bf LO}			& $-0.250$ & $0.875$ & $1.25$ & $1.44$ & $1.55$ & $1.62$ & $1.68$ & $1.72$ & $1.75$ & $1.77$\\
&{\bf NLO Pad\'e (1,1)}		& $1.32$ & $1.55$ & $1.67$ & $1.73$ & $1.78$ & $1.81$ & $1.83$ & $1.85$ & $1.87$ & $1.88$\\
&{\bf N$^2$LO Pad\'e (2,1)}     & $0.238$ & $1.17$ & $1.48$ & $1.63$ & $1.72$ & $1.78$ & $1.82$ & $1.85$ & $1.87$ & $1.89$\\[0.5em]
					& $\boldsymbol{\Delta(\Bil{3}{adj}{\mu\nu\rho})}$ &&&&&&&&&&\\
&{\bf LO}			& $2$ & $2$ & $2$ & $2$ & $2$ & $2$ & $2$ & $2$ & $2$ & $2$\\
&{\bf NLO}			& $0.500$ & $1.25$ & $1.50$ & $1.62$ & $1.70$ & $1.75$ & $1.79$ & $1.81$ & $1.83$ & $1.85$\\  
&{\bf N$^2$LO Pad\'e (2,1)}	& $1.69$ & $1.75$ & $1.79$ & $1.81$ & $1.84$ & $1.85$ & $1.87$ & $1.88$ & $1.89$ & $1.90$\\
&{\bf N$^2$LO Pad\'e (1,2)}	& $1.99$ & $1.95$ & $1.93$ & $1.92$ & $1.91$ & $1.91$ & $1.91$ & $1.92$ & $1.92$ & $1.92$\\[1em]
$\boldsymbol{\Delta(\cBil{1}{adj}{\mu})}$ & $\boldsymbol{\Delta(\Bil{2}{adj}{\mu\nu})}$ &&&&&&&&&&\\	 
&{\bf LO}			& $2.75$ & $2.38$ & $2.25$ & $2.19$ & $2.15$ & $2.12$ & $2.11$ & $2.09$ & $2.08$ & $2.08$\\
&{\bf NLO}			& $-1.58$ & $1.09$ & $1.59$ & $1.76$ & $1.85$ & $1.89$ & $1.92$ & $1.94$ & $1.95$ & $1.96$\\      
&{\bf N$^2$LO Pad\'e (2,1)}	& $1.58$ & $1.88$ & $1.95$ & $1.97$ & $1.98$ & $1.99$ & $1.99$ & $1.99$ & $1.99$ & $1.99$\\
&{\bf N$^2$LO Pad\'e (1,2)}     & $2.00$ & $2.09$ & $2.08$ & $2.06$ & $2.05$ & $2.03$ & $2.02$ & $2.02$ & $2.01$ & $2.01$\\\hline
\end{tabular}}
\caption{
$\epsilon$-expansion extrapolations of scaling dimension of the $d=3$ bilinear
operators $\cBil{0}{sing}{}$, $\cBil{0}{adj}{}$, and the conserved
current $\cBil{1}{adj}{\mu}$.
In cases in which the NLO Pad\'e (1,1) approximant is singular we 
list instead the values of the fixed-order NLO prediction.
Only three significant digits are being displayed.
\label{tab:bilinear}}
\end{table}
In figure~\ref{fig:DeltaBilinearsNftest} we plot the extrapolations
for the scaling dimension of the conserved flavor-nonsinglet current 
$\cBil{1}{adj}{\mu}$ as a function of $\Nf$.
We observe that both N$^2$LO Pad\'e approximants are closer to $2$ than the LO and NLO ones, and they remain close to $2$
even for small values of $\Nf$.
We consider this to be a successful test of the $\epsilon$-expansion, 
which supports its viability as a tool to study QED$_3$.

\begin{figure}[]
\begin{center}
\includegraphics{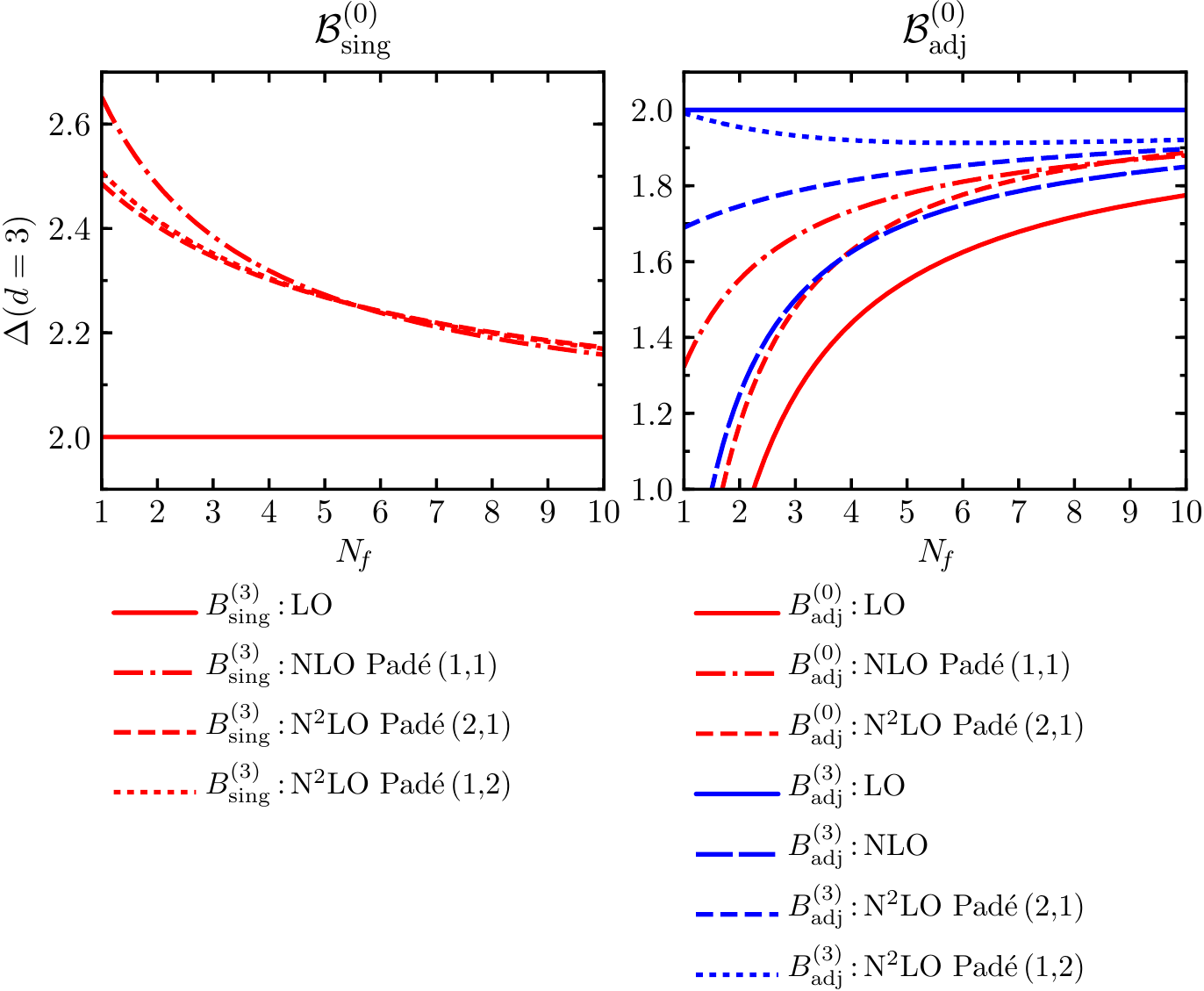}
\end{center}
\caption{$\epsilon$-expansion predictions for the scaling dimension
of the scalar bilinear operators $\cBil{0}{sing}{}$ (left panel), 
and $\cBil{0}{adj}{}$ (right panel) at $d=3$. 
The different colors for $\cBil{0}{adj}{}$ correspond to estimates from 
different continuations of the operator (see legend).
\label{fig:DeltaBilinearsNfpredictions}}
\end{figure}

In figure~\ref{fig:DeltaBilinearsNfpredictions} we plot 
the various extrapolations
for the scaling dimension of the two scalar operators
$\cBil{0}{sing}{}$ and $\cBil{0}{adj}{}$
as a function of $\Nf$.
For $\cBil{0}{sing}{}$ we find good convergence behaviour 
between the NLO Pad\'e (1,1) and the two N$^2$LO Pad\'e approximations.
Therefore, for this observable we are able to provide a rather 
convincing estimate.
We do stress, however, that the comparison of the various approximations does not provide rigorous error estimates, 
since the error due to the extrapolation is not under control. 
For $\cBil{0}{adj}{}$ we have two different operators 
that provide a continuation to $d=4-2\epsilon$. 
It is encouraging that as the order increases, the 
two resulting estimates approach each other. 
Even so, we find that for small $\Nf$ the N$^2$LO Pad\'e approximations 
are spread, so the $\epsilon$-expansion at this order does not provide a 
definite prediction. 
As $\Nf$ increases the situation improves, 
namely all NLO and N$^2$LO approximations begin to converge. 

In table \ref{tab:bilinear} we list the numerical values for the 
various estimates of the bilinear scaling dimensions 
for $\Nf = 1,\dots,10$.

Next, we compare to the large-$\Nf$ predictions for the scaling dimensions of the bilinears.
The Pad\'e approximants used to 
estimate the dimensions of $\cBil{0}{sing}{}$ and $\cBil{0}{adj}{}$ 
do not develop a pole in the extrapolation region $0\leq \epsilon \leq \frac 12$ for any value of $\Nf \geq 1$. Therefore, we can consider 
the approximants evaluated at $\epsilon = \frac 12$, 
as a function of $\Nf$, expand them around $\Nf = \infty$, i.e.,
\begin{equation}
\Delta^{\text{Pad\'e}}(k,l)\biggl\vert_{\epsilon = \frac 12} = 2 + \frac{c^{(k,l)}}{\Nf} + \order{\Nf^{-2}}~,
\end{equation} 
and compare the coefficient $c^{(k,l)}$ with its exact value 
obtained from the  
large-$\Nf$ expansion, $c^{\text{large-}\Nf}$. In what follows we use $\simeq$ to denote that we display only two significant digits.

For $\cBil{0}{sing}{}$, the prediction from large-$\Nf$ is \cite{PhysRevB.76.149906, PhysRevB.77.155105}
\begin{equation}
\cBil{0}{sing}\text{:} \quad\quad\quad c^{\text{large-}\Nf} = \frac{64}{3\pi^2} \simeq 2.2~,
\end{equation}
and the extrapolation obtained 
from the three-form singlet gives
\begin{equation}
\Bil{3}{sing}\text{:} \quad\quad c^{(1,2)} \simeq 2.3~, \quad c^{(2,1)} \simeq 2.4~.
\end{equation}

For $\cBil{0}{adj}{}$, the prediction from large-$\Nf$ is \cite{PhysRevB.72.104404, PhysRevB.77.155105}
\begin{equation}
\cBil{0}{adj}\text{:} \quad\quad\quad c^{\text{large-}\Nf} = -\frac{32}{3\pi^2} \simeq -1.1~,
\end{equation}
and the extrapolations obtained 
from the three-form and scalar adjoints give
\begin{align}
\Bil{3}{adj}\text{:} &\quad\quad c^{(1,2)} \simeq -1.4 ~, \quad c^{(2,1)} \simeq -1.4 ~,\\
\Bil{0}{adj}\text{:} &\quad\quad c^{(2,1)} = 0~.
\end{align}
This suggests that 
the extrapolation of the three-form may provide a better estimate 
for the scaling dimension of the adjoint scalar at this order.

\section{Conclusions and future directions\label{sec:conclusions}}
We employed the $\epsilon$-expansion to compute  
scaling dimensions of four-fermion and bilinear
operators at the IR fixed point of QED in $d=4-2\epsilon$.
We estimated the corresponding
value for the physically interesting case of $d=3$. 
The results seem to confirm the expectations from the 
enhancement of the global symmetry as $d\to 3$ (see
figures \ref{fig:DeltaBilinearsNftest} and \ref{fig:DeltaBilinearsNfpredictions}).
Therefore, going beyond the leading order gave us more confidence 
that the continuation is sensible.
At the same time, it appears that ---with the exception of the scalar-singlet bilinear--- 
to obtain precise estimates for the scaling dimensions for small values of $\Nf$ 
requires even higher-order computations and perhaps more 
sophisticated resummation  techniques (see for instance chapter $16$ of 
ref.~\cite{Kleinert:2001ax} and references therein).
The computation of such higher orders in $\epsilon$ via
the standard techniques used in the present work
would require hard Feynman-diagram calculations.

In recent years, several authors exploited conformal symmetry to 
introduce a variety of novel techniques to compute observables 
of the fixed point in $\epsilon$-expansion.
Ref.~\cite{Rychkov:2015naa} proposed an approach based on 
multiplet recombination, further applied and developed in refs.~\cite{Basu:2015gpa, Ghosh:2015opa, 
Raju:2015fza, Skvortsov:2015pea, Giombi:2016hkj, Nii:2016lpa, Yamaguchi:2016pbj, 
Bashmakov:2016uqk, Hasegawa:2016piv, Roumpedakis:2016qcg, Giombi:2017rhm, PhysRevLett.118.061601, Gliozzi:2017hni, Codello:2017qek, Behan:2017dwr, Behan:2017emf}.
Another approach is the analytic bootstrap, either together with  
the large-spin expansion \cite{Alday:2016jfr}, or in its Mellin-space version \cite{Sen:2015doa, Gopakumar:2016wkt, Gopakumar:2016cpb, Dey:2016mcs}.
Finally, ref.~\cite{Liendo:2017wsn} aimed at directly computing the dilatation operator 
at the Wilson--Fisher fixed point.
It would be interesting to attempt to apply these techniques to 
QED in $d=4-2\epsilon$.

On a different note, ref.~\cite{Komargodski:2017keh} recently argued
that QCD$_3$ with massless quarks 
undergoes a transition from a conformal IR phase, 
which exists for sufficiently large number of flavors,
to a symmetry-breaking phase when $\Nf \leq \Nf^c$. 
This is analogous to the long-standing conjecture for QED$_3$, 
and so
four-fermion operators may play the same role. 
Therefore, at least for the case of zero Chern--Simons level, 
$\epsilon$-expansion can be employed in a similar manner to estimate $\Nf^c$. 
A LO estimate appeared in ref.~\cite{Goldman:2016wwk}. 
In light of our results for QED$_3$, it would be worth studying 
how this estimate is modified at NLO.

\vspace{1em}
\noindent{\bf Acknowledgements:} 
we thank Joachim Brod, Martin Gorbahn, John Gracey, Igor Klebanov, Zohar Komargodski, 
and David Stone for their interest and the many helpful discussions.
We are also indebted to the Weizmann Institute of Science, in which this research
began.
Research at Perimeter Institute is supported
by the Government of Canada through Industry Canada and by the Province of Ontario
through the Ministry of Research \& Innovation.

\appendix

\section{Feynman rules\label{app:feynmanrules}}
From the QED Lagrangian in $R_\xi$-gauge,
\begin{equation}
\Lag_{\text{QED} + \text{g.f.}}  = - {1\over 4} F^{\mu\nu} F_{\mu\nu} - {1 \over 2 \xi} (\partial_\mu A^\mu)^2 +  \BPsi_a i \gamma^\mu D_\mu \Psi^a~,
\end{equation}
we obtain the  Feynman rules
\begin{align}
\parbox{40mm}{\vspace*{-1em}\includegraphics{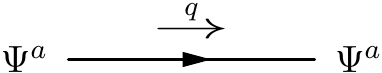}} & =\frac{i}{\slashed{q}}~,\\
\parbox{40mm}{\vspace*{-1em}\includegraphics{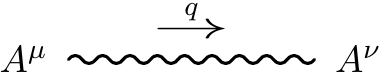}}     & =-\frac{i}{q^2}\left(\eta^{\mu\nu}-(1-\xi)\frac{q^\mu q^\nu}{q^2} \right)~, \\
\parbox{40mm}{\centering\includegraphics{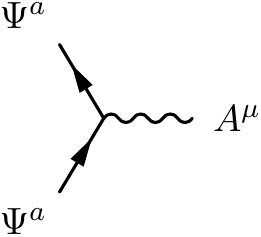}}& =-i e\gamma^\mu~. 
\label{}
\end{align}
There is one additional counterterm coupling that we need to specify.
It is a relic of the procedure with which we regulate IR divergences
(see section~\ref{sec:Loops}), which essentially breaks gauge invariance.
For this reason to consistently renormalize Green's function we need to include 
a counterterm analogous to a mass for the photon, i.e.,
\begin{align}
\parbox{40mm}{\vspace*{0em}\includegraphics{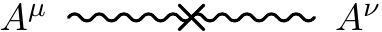}} = 
	-i \,\delta \mira^2\, \eta^{\mu\nu}\,.
\label{eq:dmira}
\end{align}
Only the one-loop value of $\delta \mira^2$ enters our
computations.
It reads
\begin{equation}
\delta \mira^2 = \alpha \frac{4\Nf}{\epsilon} \mira^2+\order{\alpha^2}\,.
\label{}
\end{equation}

To find the EOM-vanishing operators at the non-renormalizable level
we apply the EOM of the fermion and photon. They read
\begin{align} 
 \gamma^\mu D_\mu \Psi^a = 0\,,\qquad 
 D_\mu \overline{ \Psi}_a \gamma^\mu = 0\,,\qquad 
 \partial^\nu F_{\mu\nu} + e \, \BPsi_a \gamma_\mu \Psi^a = 0\,. 
\end{align}
For brevity we use the shorthand notation
$\gamma_\mu D^\mu\equiv \sD$
and use an arrow to indicate the direction in which the derivative in $\sD$ acts,
i.e.\ $\sDl$ and $\sDr\equiv \sD$.

We consider the Lagrangian with additional couplings proportional to the operators introduced in section~\ref{sec:operatorbasis} 
\begin{equation}
\Lag  = \Lag_{\text{QED}} +\sum_i C^i \cO_i~.
\end{equation}

To compute the Green's function we need the Feynman rules 
of the operators we insert, as well as all the structures 
that we need to project the amplitude. 
For instance, to renormalize the Green's function of $\BPsi\Psi A^\mu$ with
one-loop insertions of $\Q_1$ we need not only the Feynman rule of
$\Q_1$, but also the $\BPsi\Psi A^\mu$ structure of all operators 
that $\Q_1$ generates at one-loop.

In our case, the Feynman rules for the following three final states suffice:
\begin{center}
\parbox{40mm}{\includegraphics{./diagrams/PsiPsiPsiPsi}}$ = i C^\cO S^{}_\cO$
\qquad\qquad
\parbox{40mm}{\includegraphics{./diagrams/PsiPsiA}}$      = i C^\cO \tilde S^{}_\cO$ \\[2em]
\parbox{40mm}{\includegraphics{./diagrams/AA}}$           = i C^\cO \hat S^{}_\cO$ 
\end{center}
where the structures 
$S_\cO$, $\tilde S_\cO$, $\hat S_\cO$
depend on the inserted operator.
For the set of operators relevant to our computation they read
\begin{align}
&\begin{aligned}\label{eq:Str}
&{\boldsymbol{S^{}_\cO}}
 & \Q_1&:&		&2 \gamma^\mu \otimes\gamma_\mu\\
&& \Q_3&:&		&2 \gamma^{[\mu}\gamma^\nu\gamma^{\rho]}\otimes
                           \gamma_{[\mu}\gamma_\nu\gamma_{\rho]}\\
&& \E_n&:&		&2 \gamma^{[\mu_1}\cdots \gamma^{\mu_n]} \otimes
                           \gamma_{[\mu_1}\cdots \gamma_{\mu_n]}
                          + \epsilon a_{n} S^{}_{\Q_1}
			  + \epsilon b_{n} S^{}_{\Q_3}\\
&& \N_1&:&		&S^{}_{\Q_1}
\end{aligned}\\
&\begin{aligned}\label{eq:tildeStr}
&{\boldsymbol{\tilde S^{}_\cO}}
 & \N_1&:&		& 1/e\left(q^2\gamma^\mu - \slashed{q} q^\mu\right)\\
&& \N_2&:&		& \tilde S^{}_{\N_1}\\
&& \N_3&:&		&e  \left( (p+q)^2 \gamma^\mu
 -\slashed{q}\slashed{p}\gamma^\mu + 2(\slashed{q}+\slashed{p}) p^\mu\right)\\
&& \N_4&:&		&
4\slashed{q}\slashed{p}\gamma^\mu 
-4\slashed{q}p^\mu 
+2(2\slashed{p} + \slashed{q})q^\mu 
-2(2p\!\cdot\!q + q^2) \gamma^\mu\\
&& \PP&:&		 	& i/e~\mira^2
 \end{aligned}\\
&\begin{aligned}\label{eq:hatStr}
&{\boldsymbol{\hat S^{}_\cO}}
 & \N_2&:&		& \frac{2}{e^2}\left(\eta^{\mu\nu} q^4 - q^2 q^\mu q^\nu\right)\\
 \end{aligned}
\end{align}

\section{Renormalization constants \label{app:renormalizationconstants}}
In this appendix we list the mixing-renormalization constants
of four-fermion operators.
First we list the constants we need to compute the ADM of
flavor-singlet four-fermion operators, which we discussed in 
the main text, and subsequently the constants entering the 
computation of the ADM of flavor-nonsinglet four-fermion operators, which 
we discuss in appendix~\ref{app:offdiagonal}.
\subsection{Flavor-singlet four-fermion operators}
The divergent and finite pieces of the one-loop constants 
of the mixing between physical and evanescent operators
are directly related to the one-loop anomalous dimension 
of eqs.~\eqref{eq:gamma10} and \eqref{eq:gamma1min1} via
\begin{align}
\Z^{(1,1)}_{\cO\,\cO'} = \frac{1}{2} \gamma^{(1,0) }_{\cO\,\cO'}\,,&
&\Z^{(1,0)}_{\cO\,\cO'} = \frac{1}{2} \gamma^{(1,-1)}_{\cO\,\cO'}\,,
\label{}
\end{align}
with $\cO,\,\cO'$ any physical or evanescent operator from
section~\ref{sec:operatorbasis}.
To extract these constant from the $\BPsi\Psi\BPsi\Psi$ Green's function
we had to first compute the one-loop mixing of the four-fermion operators into the 
EOM-vanishing operator $\N_1$.
For the physical operators the corresponding constants are
\begin{align}
\Z^{(1,1)}_{\Q_1\,\N_1} &=  -\frac{4}{3} (2 \Nf+1)\,,&
\Z^{(1,0)}_{\Q_1\,\N_1} &= 0\,, &\\
\Z^{(1,1)}_{\Q_3\,\N_1} &=  -8\,,&
\Z^{(1,0)}_{\Q_3\,\N_1} &= 0\,, &
\label{}
\end{align}
and for the evanescent operators they are
\begin{align}
\Z^{(1,1)}_{\E_n\,\N_1} =&  0\,,&
\Z^{(1,0)}_{\E_n\,\N_1} =&  -(-1)^{\frac{n(n-1)}{2}}16(n-2) (n-5)!&\nonumber\\
&&& -\frac{4}{3} (2 \Nf+1) a_n - 8 b_n\,,
\label{}
\end{align}
with $n$ an odd integer $\ge5$.
To compute these constants for generic $n$ we used Clifford-algebra 
identities from ref.~\cite{Kennedy:1981kp}.

As explained in section~\ref{sec:greensfunctions}, in the computation of
the mixing at two-loop level more operators enter.
The only one-loop mixings entering the computation, apart from those above,
is the mixing of the physical four-fermion operators into the EOM-vanishing operator
$\N_2$, and the gauge-variant operator $\PP$. The former vanish, i.e.,
\begin{align}
\Z^{(1,1)}_{\Q\,\N_2} = 0\,,\qquad
\Z^{(1,0)}_{\Q\,\N_2} = 0\,,
\label{}
\end{align}
and the latter read
\begin{align}
\Z^{(1,1)}_{\Q_1\,\PP} = 8\Nf+4\,,\qquad
\Z^{(1,1)}_{\Q_3\,\PP} = 4\,,\qquad
\Z^{(1,0)}_{\Q\,\PP}   = 0\,,
\label{}
\end{align}
with $\Q=\Q_1,\,\Q_3$.
We do not list the corresponding constants for the evanescent operators
because they do not enter the two-loop computation of the  
mixing of physical operators.

In table~\ref{tab:greensfunctions} we summarised on which renormalization constants
the Green's functions we computed depend on.
We see that to determine the two-loop mixing of the four-fermion operators we
first need to determine the two-loop mixing of the physical operators into
the two EOM-vanishing operators $\N_1$ and $\N_2$. The corresponding constants read
\begin{align}
\Z^{(2,2)}_{\Q_1\,\N_2} =& \frac{8}{9} \Nf (2 \Nf+1)\,, &
\Z^{(2,1)}_{\Q_1\,\N_2} =& \frac{8}{9} \Nf		\,, & \\
\Z^{(2,2)}_{\Q_1\,\N_1} =& -\frac{4}{9} (12 \Nf^2+ 10 \Nf+11)\,, &
\Z^{(2,1)}_{\Q_1\,\N_1} =& \frac{4}{27}(\Nf+11)\,, & \\[1em]
\Z^{(2,2)}_{\Q_3\,\N_2} =& \frac{16}{3} \Nf	\,, &
\Z^{(2,1)}_{\Q_3\,\N_2} =& \frac{32}{9} \Nf	\,, & \\
\Z^{(2,2)}_{\Q_3\,\N_1} =& -\frac{8}{3} (24 \Nf+11)\,, &
\Z^{(2,1)}_{\Q_3\,\N_1} =& -\frac{8}{9} (103 \Nf+86) &\nonumber\\
			&&&-\frac{4}{3} a_5 (2 \Nf+1) -8 b_5    \,.&
\label{}
\end{align}

Finally, the two-loop mixing constants of the two physical operators read
\begin{align}
\Z^{(2,2)}_{\Q\,\Q'} =&
\begin{bmatrix}
 \frac{2}{9} \left(24 \Nf^2+20 \Nf+103\right) & \frac{2}{3} (3 \Nf+1) \\
 \frac{88}{3} (3 \Nf+1) & 22 \\
\end{bmatrix}\,,\\[1em]
\Z^{(2,1)}_{\Q\,\Q'} =&
\begin{bmatrix}
 -\frac{1}{54} (8 \Nf+2275) & -\frac{1}{9} (3 \Nf+49) \\
 \frac{4}{9} (107 \Nf+253) & \frac{5}{6} (8 \Nf+9) \\
\end{bmatrix}\nonumber\\
+&a_5
\begin{bmatrix}
 -\frac{1}{2} & 0 \\
 \frac{2}{3} (3 \Nf+14) & 1 \\
\end{bmatrix}
+b_5
\begin{bmatrix}
 0 & -\frac{1}{2} \\
 44 & -\frac{2}{3}  (\Nf-12) \nonumber\\
\end{bmatrix}\\
+& a_7
\begin{bmatrix}
 0 & 0 \\
 -\frac{1}{2} & 0 \\
\end{bmatrix}
+ b_7
\begin{bmatrix}
 0 & 0 \\
 0 & -\frac{1}{2} \\
\end{bmatrix}\,,
\label{}
\end{align}
with $\Q=\Q_1,\,\Q_3$.

\subsection{Flavor-nonsinglet four-fermion operators}
The renormalization of the Green's functions with insertions of 
flavor-nonsinglet four-fermion operators is analogous to the one 
with flavor-singlets but less involved.
Their flavor-off-diagonal structure forbids them to receive contributions from any 
EOM-vanishing or gauge-variant operator at two-loop order.
Therefore, in this case we only need the mixing constants 
within the physical and evanescent sectors.

As in the flavor-singlet case, the one-loop mixing is directly related 
to the one-loop anomalous dimensions of eqs.~\eqref{eq:gamma10off} and
\eqref{eq:gamma1min1off} via
\begin{align}
\Z^{(1,1)}_{\cO\,\cO'} = \frac{1}{2} \gamma^{(1,0) }_{\cO\,\cO'}\,,&
&\Z^{(1,0)}_{\cO\,\cO'} = \frac{1}{2} \gamma^{(1,-1)}_{\cO\,\cO'}\,,
\label{}
\end{align}
with $\cO,\,\cO'$ any physical or evanescent flavor-nonsinglet four-fermion
operator; the one-loop anomalous dimensions above are given
in appendix~\ref{app:offdiagonal}.
Finally, the two-loop mixing constants of the two physical operators read
\begin{align}
\Z^{(2,2)}_{\Q\,\Q'} =&
\begin{bmatrix}
 18 & \frac{2 }{3}\Nf \\
 24 \Nf & 18 \\
\end{bmatrix}\,,\\[1em]
\Z^{(2,1)}_{\Q\,\Q'} =&
\begin{bmatrix}
 -\frac{81}{2} & -\frac{1}{9} (\Nf+63) \\
 -4 (11 \Nf-9) & -\frac{1}{6} (32 \Nf+3) \\
\end{bmatrix}\nonumber\\
+&a_5
\begin{bmatrix}
 -\frac{1}{2} & 0 \\
 -\frac{2}{3} (\Nf-12) & 1 \\
\end{bmatrix}
+b_5
\begin{bmatrix}
 0 & -\frac{1}{2} \\
 36 & -\frac{2}{3} (\Nf-12) \\
\end{bmatrix}\nonumber\\
+& a_7
\begin{bmatrix}
 0 & 0 \\
 -\frac{1}{2} & 0 \\
\end{bmatrix}
+ b_7
\begin{bmatrix}
 0 & 0 \\
 0 & -\frac{1}{2} \\
\end{bmatrix}\,,
\label{}
\end{align}
with $\Q=\Q_1,\,\Q_3$.

\section{Flavor-nonsinglet four-fermion operators\label{app:offdiagonal}}
In the main part of this work we investigated bilinear and flavor-singlet 
four-fermion operators.
There exist also four-fermion operators that are not singlets under flavor.
The ones we consider in this appendix are spanned by the basis
\begin{align}
\Q_1  &=T^{ac}_{bd}	(\BPsi_a \gamma^\mu \Psi^b)
			(\BPsi_c \gamma^\mu \Psi^d)\,, \\
\Q_3  &=T^{ac}_{bd}	(\BPsi_a\Gamma^{3\,\mu\nu\rho}\Psi^b)
			(\BPsi_c\Gamma^3_{\mu\nu\rho}\Psi^d)\,,\\
\E_n  &= T^{ac}_{bd}	(\BPsi_a\Gamma^{n\,\mu_1\dots \mu_n}\Psi^b)
			(\BPsi_c\Gamma^n_{\mu_1\dots \mu_n}\Psi^d)
         + \epsilon a_n \Q_1 + \epsilon b_n \Q_3\,,
\label{eq:offQ1Q3En}
\end{align}
with $T^{ac}_{db}=T^{ca}_{bd}$ and $T^{ac}_{ad}=T^{ab}_{bd}=0$.
The computation of their ADM at one- and two-loop order entails only a subset of
the Feynman diagrams needed for flavor-singlet case and is actually less involved 
as discussed in appendix~\ref{app:renormalizationconstants}.
In this appendix we present their ADM and their scaling dimensions 
at the IR fixed point in $d=4-2\epsilon$, and use this to estimate the corresponding $d=3$ observables.

\begin{table}
\makebox[\textwidth][c]{
\begin{tabular}{lrrrrrrrrrr}
\multicolumn{1}{r}{$\boldsymbol{\Nf}$}  & $1$ & $2$ & $3$ & $4$ & $5$ & $6$ & $7$ & $8$ & $9$ & $10$ \\
\hline\hline
$\boldsymbol{(\Delta_{1})_1}$   & $-9.00$ & $-4.50$ & $-3.00$ & $-2.25$ & $-1.80$ & $-1.50$ & $-1.29$ & $-1.12$ & $-1.00$ & $-0.900$\\
$\boldsymbol{(\Delta_{2})_1}$   & $35.6$ & $8.53$ & $3.63$ & $1.95$ & $1.19$ & $0.782$ & $0.544$ & $0.393$ & $0.292$ & $0.221$\\[0.5em]
$\boldsymbol{(\Delta_{1})_2}$   & $9.00$ & $4.50$ & $3.00$ & $2.25$ & $1.80$ & $1.50$ & $1.29$ & $1.12$ & $1.00$ & $0.900$\\
$\boldsymbol{(\Delta_{2})_2}$   & $-101$ & $-29.3$ & $-14.9$ & $-9.40$ & $-6.67$ & $-5.09$ & $-4.08$ & $-3.38$ & $-2.87$ & $-2.49$\\\hline
\end{tabular}}
\caption{
Three significant digits of the one-loop, $(\Delta_{1})_i$, and the two-loop, $(\Delta_{2})_i$,
contributions to the scaling dimension of the flavor-nonsinglet four-fermion operators for
various cases of $\Nf$.
To obtain the two-loop $(\Delta_{2})_i$ values we implemented the
algorithm to include the effect of evanescent operators \cite{DiPietroStamou1}.
\label{tab:offdiagonalD2}
}
\end{table}

In the flavor scheme, the full one-loop ADM of the physical 
and evanescent operators and the two-loop entries required read:
\begin{align}
\gamma^{(1,0)}_{n m} &=
\left\{
\begin{array}[]{cl} 
 2n(n-1)(n-5)(n-6)&\text{for}\quad m=n-2\\
-4(n-1)(n-3)     &\text{for}\quad m=n\\
2               &\text{for}\quad m=n+2\\
0               &\text{otherwise\,,}\\
\end{array}
\right.\label{eq:gamma10off}\\[1em]
\gamma^{(1,-1)}_{n m} &=
\left\{
\begin{array}[]{ll}
        - 2  n(n-1)(n-5)(n-6) a_{n-2}\\
        \quad+ 4 (n-1)(n-3) a_{n}- 2 a_{n+2} + 72 b_n
        &\text{for}~m=1,\,n\ge5\\[1em]
        - 80 \delta_{n5}\\
        \quad-  2n(n-1)(n-5)(n-6)  b_{n-2}\\
        \qquad+ 4(n-1)(n-3) b_{n}- 2b_{n+2}+ 2a_n 
        &\text{for}~m=3,\,n\ge5\\[1em]
\hspace*{7em}0 &\text{otherwise\,,}
\end{array}
\right.
\label{eq:gamma1min1off}\\[1em]
\gamma^{(2,0)}_{nm} &=
\left\{
\begin{array}[]{ll}
	\begin{bmatrix}
          - 162    & -28-\frac{4}{9}\Nf \\
            144 -176 \Nf & 78- \frac{64}{3}\Nf
	\end{bmatrix}+&\\
	+ a_5 
	\begin{bmatrix}
        -2 & 0 \\
        -\frac{8}{3} \Nf & 2 \\
        \end{bmatrix}
 	+ b_5 
	\begin{bmatrix}
        0 & -2 \\
        72 & -\frac{8}{3} \Nf \\
        \end{bmatrix}
        &\text{for}~n,m=1,3\\[2em]
	\hspace*{7em}0 &\text{for}~n\ge 5\text{ and }m=1,3\\[1em]
	\hspace*{5em}\text{not required} & \text{otherwise}\,. \label{eq:gamma20qqoff}
\end{array}
\right.
\end{align}
The part of the  one-loop result that does not depend on $a_n$ and $b_n$ was first
computed in ref.~\cite{Dugan:1990df}.

Next we evaluate these ADMs at the fixed point
\begin{align}
(\gamma_1^*)_{nm} &= \frac{3}{2 \Nf}\times
\left\{
\begin{array}[]{cl}
n(n-1)(n-5)(n-6)&\text{for}\quad m=n-2\\
-2(n-1)(n-3)    &\text{for}\quad m=n\\
1               &\text{for}\quad m=n+2\\
0               & \text{otherwise\,.}
\end{array}
\right.\\[2em]
(\gamma_2^*)_{nm} &=
\left\{
\begin{array}[]{ll}
	-\frac{1}{8 \Nf^2}
	\begin{bmatrix}
	729             & 153 + 2 \Nf\\
	324 + 792 \Nf   & -351 + 96\Nf
	\end{bmatrix}   & \\
	+\frac{3}{8\Nf^2}a_5
	\begin{bmatrix}
	-3 & 0\\
	-4\Nf & 3
	\end{bmatrix}
	+\frac{3}{8\Nf^2}b_5
	\begin{bmatrix}
	0 & -3\\
	108 & -4\Nf
	\end{bmatrix}
	&\text{for}\quad n,m=1,3\,,\\[2em]
	\frac{3}{2 \Nf}  \left( -n(n-1)(n-5)(n-6) a_{n-2} \right. \\
	\left.\qquad + 2 (n-1)(n-3)a_{n} -  a_{n+2} + 36 b_n \right)
	&\text{for}~m=1,\,n\ge5~,\\[1em]
	- \frac{60}{\Nf} \delta_{n5}\\
	\quad+ \frac{3}{2 \Nf}\left( - n(n-1)(n-5)(n-6)  b_{n-2} \right.\\
	\qquad\left. + 2(n-1)(n-3) b_{n}- b_{n+2}+ a_n \right)
	&\text{for}~m=3,\,n\ge5~,\\[1em]
	\hspace*{5em}\text{not required}  & \text{otherwise\,.}
\end{array}
\right.
\end{align}
Following ref.~\cite{DiPietroStamou1} we shift to the scheme in which the 
physical--physical subblock forms an invariant subspace. In this scheme 
we are able to extract the scheme-independent $\order{\epsilon^2}$ corrections
to the scaling dimensions, i.e., the $(\Delta_2)_i$s.
In table~\ref{tab:offdiagonalD2} we list the values for the representative 
cases of $\Nf=1,\dots,10$ and in table~\ref{tab:delta12LOandNLOoff} the LO and NLO
predictions for the scaling dimensions at $d=3$.
The NLO Pad\'e (1,1) prediction of  scaling dimension $(\Delta)_2$ contains poles 
in the extrapolation region $\epsilon\in[0,\frac{1}{2}]$, so we list the fixed-order 
NLO prediction instead.

\begin{table}
\makebox[\textwidth][c]{
\begin{tabular}{llrrrrrrrrrr}
&\multicolumn{1}{r}{$\boldsymbol{\Nf}$} 		& $1$ & $2$ & $3$ & $4$ & $5$ & $6$ & $7$ & $8$ & $9$ & $10$ \\
\hline\hline
$\boldsymbol{(\Delta)_{1}}$
& {\bf LO}	           & $-0.500$ & $1.75$ & $2.50$ & $2.88$ & $3.10$ & $3.25$ & $3.36$ & $3.44$ & $3.50$ & $3.55$\\
& {\bf NLO Pad\'e (1,1)}   & $3.26$ & $3.17$ & $3.22$ & $3.30$ & $3.37$ & $3.43$ & $3.49$ & $3.53$ & $3.57$ & $3.60$\\[0.5em]
$\boldsymbol{(\Delta)_{2}}$
&{\bf LO}		  & $8.50$ & $6.25$ & $5.50$ & $5.12$ & $4.90$ & $4.75$ & $4.64$ & $4.56$ & $4.50$ & $4.45$\\
&{\bf NLO}   		  & $-16.7$ & $-1.09$ & $1.78$ & $2.78$ & $3.23$ & $3.48$ & $3.62$ & $3.72$ & $3.78$ & $3.83$\\\hline
\end{tabular}}
\caption{
LO and either NLO Pad\'e (1,1) or fixed-order NLO predictions for the 
scaling dimension of the two flavor-nonsinglet four-fermion operators at 
$d=3$ for various values of $\Nf$.
Only three significant digits are being displayed.
\label{tab:delta12LOandNLOoff}
}
\end{table}

\addcontentsline{toc}{section}{References}
\bibliographystyle{JHEP}
\bibliography{references}

\end{document}